 \documentclass[preprint2]{aastex}

\def\la{\lower.5ex\hbox{$\; \buildrel < \over \sim \;$}}
\def\ga{\lower.5ex\hbox{$\; \buildrel > \over \sim \;$}}

\shorttitle{Sustaining star formation rates in spirals}
\shortauthors{Vollmer \& Leroy}

\begin{document}


\title{Sustaining star formation rates in spiral galaxies \\
Supernova-driven turbulent accretion disk models applied to THINGS galaxies}


\author{Bernd~Vollmer}
\affil{CDS, Observatoire astronomique, UMR 7550, 11 rue de l'universit\'e, 67000 Strasbourg, France}\email{bvollmer@astro.u-strasbg.fr}

\and

\author{Adam~K.~Leroy}
\affil{National Radio Astronomy Observatory, 520 Edgemont Road,  Charlottesville, Virginia, 22903, USA}
\affil{Max Planck Institut f\"{u}r Astronomie, K\"{o}nigstuhl 17, 69117 Heidelberg, Germany}
\affil{Hubble Fellow}



\begin{abstract}
Gas disks of spiral galaxies can be described as clumpy accretion
disks without a coupling of viscosity to the actual thermal state of
the gas. The model description of a turbulent disk consisting of
emerging and spreading clumps \citep{Vollmer} contains free
parameters, which can be constrained by observations of molecular gas,
atomic gas and the star formation rate for individual galaxies. Radial
profiles of 18 nearby spiral galaxies from THINGS, HERACLES, SINGS,
and GALEX data are used to compare the observed star formation
efficiency, molecular fraction, and velocity dispersion to the model.
The observed radially decreasing velocity dispersion can be reproduced
by the model. In the framework of this model the decrease in the
inner disk is due to the stellar mass distribution which dominates the
gravitational potential.  Introducing a radial break in the star
formation efficiency into the model improves the fits
significantly.  This change in star formation regime is realized by
replacing the free fall time in the prescription of the star formation
rate with the molecule formation timescale. Depending on the star
formation prescription, the break radius is located near the
transition region between the molecular-gas-dominated and
atomic-gas-dominated parts of the galactic disk or closer to the
optical radius.  It is found that only less massive galaxies ($\log M
({\rm M}_{\odot}) \la 10)$) can balance gas loss via star formation by
radial gas accretion within the disk.  These galaxies can thus access
their gas reservoirs with large angular momentum.  On the other hand,
the star formation of massive galaxies is determined by the external
gas mass accretion rate from a putative spherical halo of ionized gas
or from satellite accretion. In the absence of this external
accretion, star formation slowly exhausts the gas within the optical
disk within the star formation timescale.

\end{abstract}





\section{Introduction}

Gas accretion plays an important role for the evolution of
galactic disks. The star formation rate in the solar neighborhood
has been approximately constant over the last $\sim 10$~Gyr
\citep{Binney2000}, while the local gas depletion time is
  $\approx 2$~Gyr \citep{Evans2008}. This suggests that the gas
consumed by star formation must be replenished via external or
internal accretion. Continuous addition of metal-poor gas would
also explain the discrepancy between the observed stellar metallicity
distribution in the solar neighborhood and that predicted by
closed-box models of chemical evolution \citep{Tinsley1981}.

The need for accretion may also be seen from observations of other
galaxies. Large spiral galaxies like the Milky Way typically have star
formation rates of a few M$_{\odot}$/yr and gas reservoirs of
  order $10^9$~M$_\odot$ in the star-forming part of the galaxy. Over
these regions, gas depletion times are 1--2 Gyr \citep{WongBlitz}, so
that for these galaxies to survive in their present configuration,
roughly the same amount of mass consumed by star formation must arrive
at the galaxy center via some type of accretion.  This accretion can
be external, infalling gas from the galactic halo, or internal, radial
gas transport via spiral arms and gas viscosity.  In the absence
  of external accretion the spiral galaxy might deplete its gas and
  become a lenticular galaxy.

Turbulent galactic disks have been studied numerically and
  analytically in recent years. \citet{WadaNorman2007} used 3D, global
  hydrodynamic simulations of the central part of a late type galaxy
  to study the density distribution of the inhomogeneous ISM. They
  found that the density distribution is well fitted by a single
  lognormal function over a wide density range. \citet{Elmegreen2002}
  first noticed that the Schmidt-Kennicutt law of the star formation
  rate can be reproduced if star formation occurs in dense gas above a
  critical density.  The simulations of \citet{WadaNorman2007} did not
  include feedback from supernova (SN) explosions.

In a series of articles,
\citet{KimOstriker}, \citet{ShettyOstriker}, and \citet{KoyamaOstriker}
studied the 2D evolution of an inhomogeneous ISM in a galactic disk
without SN feedback.

\citet{KimOstriker} emphasize the importance of gravity due to gas
and stars for driving interstellar turbulence. When strong feedback from
SN is included \citep{ShettyOstriker}, the fraction of dense gas
and thus star formation is reduced compared to the models of 
\citet{KimOstriker} and turbulence is driven by expanding shells
in overdense regions. On the other hand, \citet{Agertz}, who
studied gaseous galactic disks by means of 3D, high-resolution hydrodynamical
simulations with and without SN feedback, found that SN feedback is an
important driver of turbulence in galaxies with star formation rates
$\geq 10^{-3}$~M$_{\odot}$yr$^{-1}$kpc$^{-2}$. This star formation
rate is found to galactic radii $\sim R_{25}$ in most of the nearby
spiral galaxies investigated by \citet{Leroy}.

\citet{ShettyOstriker} note that the gas
disk thickness is important for setting the index of the
Schmidt-Kennicutt law. \citet{KoyamaOstriker} studied the relationships
among pressure, vertical distribution of gas, and the fraction of
dense gas in 2D hydrodynamic simulations of a starforming galactic
disk with SN feedback. They found that vertical hydrostatic equilibrium
gives a good estimate for the mean midplane pressure of the ISM.
In their models they recover the observed ratio $R_{\rm mol}=M_{\rm H_2}/M_{\rm HI}$
if the gas surface density is proportional to the epicyclic frequency.
\citet{KoyamaOstriker} state that the empirical result that $R_{\rm mol}$ is proportional 
to the mean midplane pressure of the ISM implies that the epicyclic frequency,
the gas surface density, and the star formation rate per unit volume
are interdependent, most probably due to a galaxy evolution with the Toomre
parameter $Q$ close to unity. If gravitational instability of a clumpy
and turbulent gas disk is taken into account, turbulence drives
the disc to a regime of transition between instability at small
scales and stability in the classical sense (Toomre $Q \sim 1$; \citet{Romeo}).

Most recently, \citet{KrumholzBurkert} have developed an analytical model
for the evolution of a thin disk of gas and stars with an arbitrary 
rotation curve that is kept in a state of marginal gravitational instability ($Q \sim 1$)
and energy equilibrium due to the balance between energy released by
accretion and energy lost due to decay of turbulence. Equilibrium disks
of this kind have been investigated by \citet{VollmerBeckert}.
\citet{KrumholzBurkert} showed that disks initially out of equilibrium
evolve into it on timescales comparable to the orbital period if the
external gas accretion rate is high.

In this article we use the analytical models of \citet[][hereafter
  VB03]{Vollmer} to estimate the radial mass accretion rate of a
sample of nearby spiral galaxies. The VB03 model treats galaxies as
clumpy accretion disks using a simplified description of turbulence
driven by supernova (SN) explosions. Because we focus on
  modeling the galactic disk within the optical radius, we neglect
  gravitational instabilities as energy source for turbulence
  \citep[see][]{Agertz}. A gas disk without star formation might
  generate turbulence via gravitational instabilities, but once stars
  form, we assume that SN dominate energy injection.  The formalism of
  our equilibrium model is close to that of
  \citet{KrumholzBurkert}. The model naturally links the physical
  properties of the galactic gas disk (surface density, molecular
  fraction, star formation rate) to the midplane pressure which mainly
  depends on the stellar surface density, epicyclic frequency, the Toomre parameter, 
  and the gas mass accretion rate \citep[similar to][]{KoyamaOstriker}.
  Furthermore, we assume the classical scale-independent Toomre
  stability criterion.

We compare these models to multiwavelength observations of 18 spiral
galaxies from the THINGS sample \citep{Walter} compiled by
\citet{Leroy}. From the comparison, we estimate the radial mass
accretion rate of each galaxy.

The VB03 model considers the gas disk of a galaxy as a single
  turbulent medium. The dissipation timescale of turbulent kinetic
energy on galactic scales is about the crossing time
\citep{Elmegreen2000}, so that for a typical driving scale length of
$\sim 100$~pc of the turbulent flow the dissipation time $\tau_{\rm d}
\sim 10$~Myr . Therefore in order to maintain turbulence, an
efficient, continuous, driving mechanism is needed. Possible
candidates for such a driving mechanism are:
\begin{itemize}
\item gravitational instabilities like spiral arms 
\citep[see, e.g.][]{GomezCox} or rotational shear 
\citep[see, e.g.,][]{WadaNorman,VollmerBeckert,KrumholzBurkert},
\item magneto-rotational instabilities \citep{BalbusHawley},
\item protostellar outflows \citep[e.g.,][]{Wolf-Chase},
\item stellar winds \citep[e.g.,][]{Vink},
\item UV radiation,
\item supernova (SN) explosions \citep[see e.g.][]{Vollmer,Agertz},
\item external accretion \citep{ElmegreenBurkert,KlessenHennebelle}.
\end{itemize}
Of these only SN explosions can balance the energy loss due to
turbulent energy dissipation if the driving length scale is $\sim
100$~pc \citep{MacLow} and the star formation rate is
$\geq 10^{-3}$~M$_{\odot}$yr$^{-1}$kpc$^{-2}$ \citep{Agertz}.
In cases where the driving length scale is of
the order of the disk thickness ($\sim 500$~pc), the energy input due
to gravitational instabilities can maintain turbulence in the
interstellar medium (ISM) \citep{VollmerBeckert,KrumholzBurkert}. 
The VB03 model and this paper consider only turbulence driven by SN.

In the VB03 model, SN-driven turbulence sets the disk structure
  and the disk structure determines the star formation rate, which in
  turn determines the SN rate. As a result, the conversion of gas
into stars is of prime importance to the model. Despite a large
variety of observations in different wavelengths, star formation in
galaxies is not well understood \citep[for recent reviews
  see][]{MacLow,McKee}. ISM turbulence in turn probably controls
  the star formation process to some degree \citep{MacLow}, but large
  scale gravitational instabilities \citep{Toomre}, thermal
  instabilities \citep{ElmegreenParravano,Wolfire}, and H$_2$
  formation \citep{Krumholz09} may all be important.

In the last decades it was thought that star
formation is controlled by the interplay between self-gravitation and
magnetic fields. However, the interstellar medium (ISM) is of
turbulent nature. Its structure is usually described as hierarchical
\citep{Scalo} over length scales of several magnitudes up to
$\sim$100~pc. The neutral phase of the ISM is not uniform but has a
fractal structure \citep{Elmegreen}.

\citet{Vollmer} developed an analytical model for
clumpy accretion disks and included a simplified description of
turbulence in the disk. In contrast to classical accretion disk theory
\citep[e.g.,][]{Pringle}, VB03 do not use the ``thermostat''
mechanism, which implies a direct coupling between the heat produced
by viscous friction and the viscosity itself, which is assumed to be
proportional to the thermal sound speed. Thus, the viscosity, which is
responsible for the gas heating, depends itself on the gas
temperature.  This leads to an equilibrium corresponding to a
thermostat mechanism.  Instead, VB03 use energy flux conservation,
where the energy flux provided by SNe is transferred through a
turbulent cascade to smaller scales where it is dissipated.  The SN
energy flux is expected to be proportional to the local star formation
rate.  In particular, the star formation recipe takes the turbulent
nature of the ISM into account. VB03 showed that it is possible to
reproduce the the local and global gas properties of the Galaxy with
this model.  It was, however, not possible at that time to compare
observed radial profile to the VB03 model.

Here, we extend this comparison to a sample of 18 nearby spiral
galaxies.  \citet{Leroy} presented radial profiles of the star
formation rate and gas surface density (CO and H{\sc i}) of these
galaxies to study their star formation efficiencies. They showed that
the SFE of H$_{2}$ alone is nearly constant (H$_{2}$ depletion time of
$1.9$~Gyr) at their 800~pc resolution. Where the interstellar medium
(ISM) is mostly H{\sc i}, however, the SFE decreases with increasing
radius, a decline reasonably described by an exponential with scale
length 0.2R$_{25}$--0.25R$_{25}$.

In this article we extend the formalism of VB03 (Sec.~\ref{sec:model})
by adding a break in the star formation rate (Sec.~\ref{sec:break}).
Beyond this break the relevant timescale for star formation is no
longer the free fall time of the most massive self-gravitating cloud
at a given galactocentric radius, but the molecular formation
timescale.  In Sec.~\ref{sec:results} and \ref{sec:discussion} the radial profiles presented in
\citet{Leroy} are compared and discussed to the extended VB03 model.  We discuss the
implications of our findings in Sec.~\ref{sec:how} and give our
conclusions in Sec.~\ref{sec:conclusions}.

\section{The analytical model \label{sec:model}}

Since the model is described in detail in VB03, we summarize here only the
basic idea and the resulting expressions for the relevant disk properties.  

The model considers the warm, cold, and molecular phases of the ISM as a
single, turbulent gas.  We assume this gas to be in vertical hydrostatic
equilibrium, with the midplane pressure balancing the weight of the gas and stellar disk.
The gas is taken to be clumpy, so that the local density is enhanced relative
to the average density of the disk. Using this local density, we calculate two
timescales relevant to star formation: the free fall of time of an individual
clump and the characteristic timescale for H$_2$ to form on grains. The longer
of these (modulo an empirical scaling factor) is taken as the governing
timescale for star formation. The star formation rate is used to calculate the
rate of energy injection by supernovae. This rate is related to the turbulent
velocity dispersion (an observable) and the driving scale of turbulence. These quantities
in turn provide estimates of the clumpiness of gas in the disk (i.e., the
contrast between local and average density) and the rate at which viscosity
moves matter inward.

The model relies on several empirical calibrations: e.g., the relationship
between star formation rate and energy injected into the ISM by supernovae,
the H$_2$ formation timescale (and its dependence on metallicity), and the
turbulent dimension of the ISM (used to relate the driving length scale to the
characteristic cloud size modulo a free parameter). As far as possible, these
are drawn from observations of the Milky Way.

We are left with three free parameters. First, there is an unknown scaling
factor relating the driving length of turbulence to the size of
gravitationally bound clumps, which we call $\delta$. Second, the point at
which the star formation timescale transitions from the free fall time to the
H$_2$ formation time is {\em a priori} unknown. Third, the mass accretion
rate, $\dot{M}$, which is related to the driving length and turbulent velocity,
is a free parameter. In Section \ref{sec:results} via comparison to the
present sample of galaxies, we constrain $\delta$.
In practice, we also treat the Toomre $Q$ parameter of the gas as a ``semi-free''
parameter, allowing it to change somewhat from the observed value.

With these assumptions in hand, we may fit the model to a galaxy by comparing
the observed kinematics, {\sc Hi}, CO, and star formation rate profiles to
those predicted by the model. The resulting fit yields the disk mass accretion
rate and several other parameters that may be checked against expectations:
the free fall time (or density) for the largest self-gravitating structures,
the driving length of turbulence, and the approximate velocity of radial
inflow.

In the remainder of this section, we discuss our assumptions in slightly more
detail, justify them via comparison to observation and theory, and note the
physics that we neglect.

{\em ISM:} Following, e.g., \citet{MacLow}, we view the ISM as a single
turbulent gas. In this picture, the warm, cold, and molecular phases of the
ISM are a single entity. Locally, the exact phase of the gas may depend on the
local pressure, metallicity, stellar radiation field, stellar winds, and
shocks. Here we view these factors as secondary, making a few simplifying
assumptions.
The equilibrium between the different phases of the ISM 
and the equilibrium between turbulence and star formation depends 
on three local timescales:
the turbulent crossing time $t_{\rm turb}^{l}$, the molecule formation timescale $t_{\rm mol}^{l}$, 
and the local free fall timescale $t_{\rm ff}^{l}$ of a cloud.

{\em Supernova-driven Turbulence:} First, we assume that the gas is turbulent,
so that the turbulent velocity is the relevant one throughout the disk 
(making the exact temperature of the gas is largely irrelevant). We assume that
this turbulence is driven by SNe and that they input their energy in turbulent eddies
that have a characteristic length scale, $l_{\rm driv}$, and a characteristic
velocity, $v_{\rm turb}$.  This driving length scale may be the characteristic
length scale of a SN bubble, but it does not have to be so. It may be set by
the interaction of multiple SN bubbles or of a SN with the surrounding ISM. We
note that based on simulations, the assumption of a single driving scale may
be a simplification \citep{JoungMacLow}. The VB03 model does not address the
spatial inhomogeneity of the turbulent driving nor the mechanics of turbulent
driving and dissipation. It is assumed that the energy input rate into the ISM due
to SNe, $\dot{E}_{\rm SN}$, is cascaded to smaller scales without losses by turbulence.
At scales smaller than the size of the largest selfgravitating clouds the energy is dissipated 
via cloud contraction and star formation. We refer to \citet{MacLow} for a review of these topics.
We limit our analytical model to the first
energy sink which is the scale where the clouds become selfgravitating. 

We can connect the energy input into the ISM by SNe directly to the star
formation rate. With the assumption of a constant initial mass function (IMF)
independent of environment one can write

\begin{equation}
  \label{eq:energyflux}
  \frac{\dot{E}_{\rm SN}}{\Delta A}=\xi\,\dot{\Sigma}_{*} 
  = \xi\,\dot{\rho}_{*} l_{\rm driv}=\Sigma \nu \frac{v_{\rm turb}^{2}}{l_{\rm driv}^{2}}\ ,
\end{equation}

where $\Delta A$ is the unit surface element of the disk.  The factor
of proportionality $\xi$ relates the local SN energy input to the
local star formation rate and is assumed to be independent of local
conditions.  $\xi$ is normalized using Galactic observations by
integrating over the Galactic disk and results in $\xi=4.6 \times
10^{-8}$~(pc/yr)$^{2}$ (see VB03).  The adopted energy that is
  injected into the ISM is $E^{\rm kin}_{\rm SN}=10^{50}$~ergs based
  on numerical studies by \citet{Thornton}.  The final two parts of
Equation \ref{eq:energyflux} assume that stars form over a
characteristic scale equal to the driving length and equate energy
output from SNe with the energy transported by turbulence (see VB03).

{\em Star Formation in Molecular Clouds:} Second, we assume that stars form
out of gravitationally bound clouds. Where the timescale to form
H$_2$ is short, we take the local gravitational free fall time, given by

\begin{equation}
  t^{\rm l}_{\rm ff}=\sqrt{\frac{3\pi}{32G\rho_{\rm cl}}}\ ,
  \label{eq:localff}
\end{equation}

\noindent to be the relevant timescale for star formation. Here $G$ is the
gravitational constant and $\rho_{\rm cl}$ the density of a single cloud.

Cloud collapse, and thus star formation can only proceed if enough molecules
form during the cloud collapse to allow the gas to continue cooling.
Therefore, where the timescale for H$_2$ formation is long (compared to the
free fall time), we view this as the relevant timescale for star formation.
We take the characteristic time to form H$_2$ out of H to be approximately

\begin{equation} \label{eq:molform}
t_{\rm mol}^{l}=\alpha / \rho_{\rm cl}\ ,
\end{equation} 
where $\alpha$ is a coefficient that depends on metallicity and
  temperature \citep{DraineBertoldi1996} and $\rho_{\rm cl}$ is the
  density of a single cloud. Unlike some other recent numerical
  and theoretical treatments of H$_2$ abundance by \citet{Robertson08}
  and \citet{Krumholz09}, we take no account of the rate of
destruction of H$_2$ via the UV radiation field. Instead, we
  assume H$_2$ to be destroyed via the collapse and subsequent star
  formation of a self-gravitating cloud.

The coefficient of the molecular formation timescale $\alpha$ is
assumed to be metallicity dependent \citep{TielensH}. Because we
  admit external gas accretion, the metallicity of the star-forming
  ISM mainly depends on the ratio of accretion to star formation rate
  $a$.  Small $a<1$ lead to a metallicity derived from a closed box
  model, whereas in the case of $a>1$ the metallicity equals the true
  stellar yield $y_{\rm true}$ \citep{Koeppen}. For gas fractions
  higher than 0.04 the difference between the two solutions is less
  than a factor of two.  Moreover, \citet{Dalcanton} showed that the
  effective yield $y_{\rm eff}=Z_{\rm gas}/\ln(1/f_{\rm gas})$, where
  $Z_{\rm gas}$ is the gas metallicity and $f_{\rm gas}$ the gas
  fraction, for disk galaxies with a rotation velocity higher than
  $100$~km\,s$^{-1}$ is approximately constant, i.e., for these
  galaxies a closed box model can be applied.  We thus feel confident
  to estimate the metallicity based on a closed box model using the
gas fraction
\footnote{Only three galaxies, IC~2574, NGC~4214, and NGC~2976, have rotation
velocities smaller than $100$~km\,s$^{-1}$. We calculated the models for
these galaxies using up to 3 times smaller effective yields. The
results where not significantly different compared to the models
based on Eq.~\ref{eq:alphacb} with the exception that the molecular
fraction decreased by the same factor as the effective yield.}:
\begin{equation} \label{eq:alphacb}
\alpha=7.2 \times 10^{7} \times \big( \log(\frac{\Sigma_{*}+\Sigma}{\Sigma})
\big)^{-1}\ {\rm yr\,M_{\odot}pc^{-3}},
\end{equation}
where $\Sigma_{*}$ is the stellar surface density and $\log$
  refers to a natural logarithm.  The calibration of the
  effective yield is based on the comparison of the observed O/H
profiles with those of the closed box model (Sect.~\ref{sec:results}).
Adopting a stellar and gas surface density of
$\Sigma_{*}=40$~M$_{\odot}$pc$^{-2}$ and $\Sigma_{\rm
  gas}=10\,$~M$_{\odot}$pc$^{-2}$ at the solar radius of the Galaxy
\citep{Allen} yields $\alpha_{\odot}=4.5 \times
10^{7}$~yr\,M$_{\odot}$pc$^{-3}$, which is twice the value used by
\citet{Hollenbach}.

This approach assumes that molecular clouds are relatively
short lived, appearing and disappearing over roughly a free fall time
(equivalently, by our construction, a turbulent crossing time); otherwise they
might reach chemical equilibrium even when the H$_2$ formation time is long
compared to the free fall time. Accordingly, we estimate the molecular
fraction in the disk from the ratio of a cloud lifetime (i.e., the crossing or
free-fall time) to the H$_2$ formation time scale:
$f_{\rm mol}=\frac{\Sigma_{\rm H_{2}}}{\Sigma_{\rm HI}+\Sigma_{\rm H_{2}}}=t_{\rm turb}^{l}/t_{\rm mol}^{l}$.

Note that the molecular formation timescale, in particular, is fairly
approximate --- as noted above we have neglected radiation field
  and temperature dependences. Therefore we allow an extra factor
$\gamma$ when comparing the two to derive the relevant timescale, with
the transition between the two regimes at $t_{\rm ff}^l = \gamma
t_{\rm mol}^l$. We derive a value of $\gamma = 0.4 \pm 0.3$ for the
VB03 star formation prescription and $\gamma = 0.12 \pm 0.06$ for the
star formation prescription following \citet{Krumholz}.  comparing the
model to the whole set of observations in Section~\ref{sec:results}.

{\em Vertical Disk Structure:} In the model, the disk scale height is
determined unambiguously by the assumption of hydrostatic equilibrium and the turbulent pressure
\citep{Elmegreen89}:

\begin{equation}
p_{\rm turb}=\rho v_{\rm  turb}^{2} = \frac{\pi}{2} G \Sigma ( \Sigma + \Sigma_{*} \frac{v_{\rm turb}}{v_{\rm disp}^{*}})~,
\label{eq:pressure}
\end{equation}

\noindent where $\rho$ is the average density, $v_{\rm turb}$ the gas turbulent
velocity in the disk, $v_{\rm disp}^{*}$ the stellar vertical velocity dispersion, 
and $\Sigma$ the surface 
density of gas and stars. The stellar velocity dispersion is calculated
by $v_{\rm disp}^{*}=\sqrt{2 \pi G \Sigma_{*} H_{*}}$, where
the stellar vertical height is taken to be $H_{*}=l_{*}/7.3$ with  $l_{*}$ being the
stellar radial scale length \citep{Kregel}. 
We neglect thermal, cosmic ray, and 
magnetic pressure. During the fitting procedure we realized that the local
pressure equilibrium is of prime importance for the goodness of the fits.

{\em Treatment as an Accretion Disk:} The turbulent motion of clouds is
expected to redistribute angular momentum in the gas disk like an effective
viscosity would do. This allows accretion of gas towards the center and makes
it possible to treat the disk as an accretion disk \citep[e.g.,][]{Pringle}.
This gaseous turbulent accretion disk rotates in a given gravitational
potential $\Phi$ with an angular velocity $\Omega=\sqrt{R^{-1}\frac{{\rm
      d}\Phi}{{\rm d}R}}$, where $R$ is the disk radius. The disk has an
effective turbulent viscosity that is responsible for mass accretion and
outward angular momentum transport. In this case, the turbulent velocity is
driven by SN explosions, which stir the disk and lead to viscous transport of
angular momentum. In addition, star formation removes gas from the viscous evolution.
Following \citet{LinPringle}, the evolution of the gas surface density 
is given by
\begin{equation}
\frac{\partial \Sigma}{\partial t}=-\frac{1}{R}\frac{\partial}{\partial R}\left(
\frac{(\partial/\partial R)[\nu \Sigma R^3 ({\rm d}\Omega/{\rm }dR)]}{({\rm d}/{\rm d}R)(R^2 \Omega)}\right)
-\dot{\Sigma}_{*}+\dot{\Sigma}_{\rm ext}\ ,
\label{eq:linpringle}
\end{equation}
where $\nu$ is the gas disk viscosity, $\Omega$ the angular velocity, 
and $\dot{\Sigma}_{\rm ext}$ is the external mass
accretion rate. In contrast to \citet{LinPringle} we assume a continuous and non-zero external gas mass 
accretion rate.
By approximating $\partial/\partial R \sim 1/R$,
the global viscous evolution becomes
\begin{equation}
\frac{\partial \Sigma}{\partial t} \sim \frac{\Sigma \nu}{R^2}-\dot{\Sigma}_{*}+\dot{\Sigma}_{\rm ext}\ .
\label{eq:parsigma}
\end{equation}
If the external and disk mass accretion rate keeps the combined Toomre parameter of the gas and stars 
smoothed over a few rotation periods to $Q_{\rm tot} \sim 1$, 
the gas surface density will only vary slowly with changes in the dark halo
mass distribution (via $\Omega$) and the stellar disk
structure, the gas loss due to star formation is balanced by external accretion
as suggested by \citet{Fraternali} and \citet{Marinacci},
and the gas disk can be regarded as being stationary ($\partial \Sigma/ \partial t = 0$). 
Indeed, local spiral galaxies
show a $Q_{\rm tot}$ not too far away from unity ($1.3$-$2.5$; \citet{Leroy}).
For such a stationary gas disk, where star formation is balanced by
external accretion, the local mass and momentum conservations yield:
\begin{equation}
\nu \Sigma=\frac{\dot{M}}{2\pi}\ ,
\label{eq:transport}
\end{equation}
where $\dot{M}$ is the mass accretion rate within the disk.
In the absence of external mass accretion,
the gas disk can be assumed to be stationary as long as the star formation timescale $t_*$ exceeds the 
viscous timescale $t_{\nu}=R^2/\nu$.
For $\dot{\Sigma}_{\rm ext} < \dot{\Sigma}_{*}$ and $t_* < t_{\nu}$ Eq.~\ref{eq:transport}
is not valid. In this case the gas disk
is rapidly turned into stars within the gas consumption time ($2$~Gyr, \citet{Evans2008}).
Since most spiral galaxies still have a significant amount of gas, we think that 
spiral galaxies are generally not in this state.
Solving the time dependent Eq.~\ref{eq:linpringle} is beyond the scope of this 
work and we apply Eq.~\ref{eq:transport}.
Radial integration of Eq.~\ref{eq:parsigma} then gives
\begin{equation}
\frac{\partial M}{\partial t} \sim \dot{M}-\dot{M}_{*}+\dot{M}_{\rm ext}\ ,
\end{equation}
where $M$ is the disk gas mass.



The viscosity is related to the driving length scale and
characteristic velocity of the SN-driven turbulence by $\nu=v_{\rm turb}l_{\rm
  driv}$ (VB03). Because the lifetime of a collapsing and starforming cloud
($t_{\rm ff}^{l} < t_{\rm turb}^{l}$) is smaller than the turnover time of the
large-scale eddy ($l_{\rm driv}/v_{\rm turb}$), the turbulent and clumpy ISM
can be treated as one entity for the viscosity description.

{\em Clumpiness:} A critical factor in the model is the relationship between
the density of individual clouds, $\rho_{\rm cl}$, and the average density of
the disk, $\rho$. It is the density of individual clouds that is relevant to
the timescale for star formation. In this model, the two are related by the
volume filling factor, $\Phi_{\rm V}$ so that $\rho_{\rm cl}=\Phi_{\rm
  V}^{-1}\rho$.

Here $\rho_{\rm cl}$ refers to the density of the largest
self-gravitating structures in the disk, so that for these structures
the turbulent crossing time and gravitational free fall time are
equal. The scale of such a cloud, $l_{\rm cl}$ is smaller than the
driving length scale $l_{\rm driv}$ by a factor $\delta$, which we do
not know {\em a priori}.

Shear due to differential galactic rotation could stabilize
  clouds, modifying the timescale for collapse. However, this effect
  is mainly important when the ratio of the cloud to disk surface
  density is lower than the ratio of cloud to disk velocity
  dispersion, which is not the case over most of the disk in a typical
  spiral. Typical GMC surface densities are $\sim
  200$~M$_{\odot}$pc$^{-2}$ \citep{Solomon}, whereas disk surface
  densities only exceed $100$~M$_{\odot}$pc$^{-2}$ in the very center
  of spiral galaxies \citep{Leroy}. The ratio of velocity dispersion
  being about $0.5$, clouds at galactic radii larger than a few kpc
  are not stabilized by shear from galactic rotation.

We can calculate the turbulent timescale for the cloud, $t_{\rm turb}^l$, for
a fractal ISM:

\begin{equation}
t_{\rm turb}^{l}=\delta^{-\frac{2}{3}-\frac{3-D}{3}}
\,l_{\rm driv}/v_{\rm turb}\ ,
\end{equation}

where $D$ is the fractal dimension \citep[see, e.g.,][]{Frisch} of the ISM.
We assume $D=2$ for a compressible, self-gravitating fluid, which is close to
the findings of \citet{Elmegreen}. Once $\delta$ and thus $t_{\rm turb}^l$ are
specified, we can solve for the density of the corresponding scale by setting
$t_{\rm ff}^l = t_{\rm turb}^l$. The volume filling factor is then defined by
comparing $\rho_{\rm cl}$ and $\rho$.

Because the relationship between the driving length scale, $l_{\rm driv}$ and
$l_{\rm cl}$ is not known beforehand, we treat this as a free parameter.
Once the volume filling factor is known (from $\delta$ or $l_{\rm cl}$), we
can calculate the local star formation rate, $\dot{\rho_*}$, via

\begin{equation}\label{eq:starform}
\dot{\rho}_{*} = \eta \frac{\rho}{t_{\rm sf}^{\rm l}}\ ,
\end{equation}

\noindent where $t_{\rm sf}^l$ is the local timescale for star
formation, either the free fall or molecular formation timescale,
depending on local conditions. We assume $\eta=7\times 10^{-3}$ in
consistency with \citet{Krumholz}.  Since in our model the
  lifetime of a cloud is the free-fall as suggested by
  \citet{Ballesteros} or the molecule formation timescale, this
  implies that during the cloud lifetime about $1$\,\% of the cloud
  mass turns into stars.

In VB03 a different star formation prescription is used:

\begin{equation}\label{eq:starformm}
\dot{\rho}_{*} = \Phi_{\rm V} \frac{\rho}{t_{\rm sf}^{\rm l}}\ .
\end{equation}

\noindent The vertically integrated star formation rate in the inner disk 
where $t_{\rm sf}^{\rm l}=t_{\rm ff}^{\rm l}=t_{\rm turb}^{\rm l}=\delta^{-1} t_{\rm turb}$ is 

\begin{equation}\label{eq:starformm1}
\dot{\Sigma}_{*} = \Phi_{\rm V} \frac{\rho}{t_{\rm ff}^{\rm l}} l_{\rm driv} = 
\delta \Phi_{\rm V} \rho v_{\rm turb}\ ,
\end{equation}

\noindent i.e. it is the mass flux density of the turbulent ISM into the regions
of star formation. This alternative description
yields fitting results that are as good
as those based on Eq.~\ref{eq:starform} (in practice $\eta \sim \Phi_{\rm V} \sim 10^{-2}$).

\subsection{A Two Part Model \label{sec:break}}

VB03 assumed that $t_{\rm sf}^l$ in Equation \ref{eq:starform} is always the
local free fall time. As described above, here we consider two regimes,
adopting instead the H$_2$ formation time when this becomes long compared to
$t_{\rm ff}^l$. Therefore, the present model has two parts. We label the
first, where $t_{\rm sf}^l = t_{\rm ff}^l$ the {\em inner disk}. We label the
second, where $t_{\rm sf}^l = t_{\rm mol}^l$ the {\em outer disk}.
The change in regime should occur at the radius where $t_{\rm ff}^l = \gamma t_{\rm mol}^l$.
Here $\gamma$ is a (free) scaling parameter that reflects a level of
uncertainty in the application of the molecular formation timescale.

An alternative interpretation of the change in star formation regime is
that (i) the star formation efficiency per free fall time or (ii) the volume or area 
filling factor of star forming regions change
from the inner to the outer disk. The latter possibility is motivated by the work of
\citet{Bush} who show that gaseous spiral
arms formed in the inner disk region can propagate into the outer gas disk
creating overdensities where stars can form. Since these gaseous arms occupy 
a smaller fraction of the disk area with increasing galactocentric radius,
the azimuthally averaged surface and area filling factors decrease.
Having this picture in mind, one can write for the outer disk

\begin{equation}
\dot{\rho}_{*}=\eta \frac{\rho}{t_{\rm mol}^{\rm l}}=\eta f_{\rm mol}  \frac{\rho}{t_{\rm ff}^{\rm l}}
=\tilde{\eta} \frac{\rho}{t_{\rm ff}^{\rm l}}
\end{equation}
and
\begin{equation}
\dot{\rho}_{*}=\Phi_{\rm V} \frac{\rho}{t_{\rm mol}^{\rm l}}=\Phi_{\rm V} f_{\rm mol}  \frac{\rho}{t_{\rm ff}^{\rm l}}
=\tilde{\Phi}_{\rm V} \frac{\rho}{t_{\rm ff}^{\rm l}}\ .
\end{equation}

\noindent In this formulation the relevant local star formation timescale is still
the local free fall time of the clouds.

\subsection{Model calculations \label{sec:calc}}

The VB03 model yields the following system of equations to describe a turbulent clumpy galactic accretion disk:
\begin{displaymath}
\nu  =  v_{\rm turb} l_{\rm driv}\ ,
\end{displaymath}
\begin{displaymath}
\nu \Sigma  =  \frac{\dot{M}}{2\pi}\ ,
\end{displaymath}
\begin{displaymath}
\Sigma  =  \rho\,H\ ,
\end{displaymath}
\begin{displaymath}
p_{\rm turb}=\rho v_{\rm  turb}^{2} = \frac{\pi}{2} G \Sigma ( \Sigma + \Sigma_{*} \frac{v_{\rm turb}}{v_{\rm disp}^{*}})~,
\end{displaymath}
\begin{displaymath}
Q  =  \frac{v_{\rm turb} \Omega}{\pi G \Sigma}\ ,
\end{displaymath}
\begin{displaymath}
\Sigma \nu \frac{v_{\rm turb}^{2}}{l_{\rm driv}^{2}} =  \xi\,\dot{\Sigma}_{*}\ ,
\end{displaymath}
\begin{displaymath}
\dot{\Sigma}_{*} =  \Phi_{\rm V} \frac{\rho}{t_{\rm SF}^{\rm l}} l_{\rm driv}\ {\rm (VB03)\ or}\ =\eta \frac{\rho}{{t_{\rm SF}^{\rm l}}} l_{\rm driv}\ {\rm (KM05)}\ ,
\end{displaymath}
\begin{displaymath}
t_{\rm SF}^{\rm l}=\sqrt{\frac{3\pi}{32G\rho_{\rm cl}}}\ {\rm or}\ =\frac{\alpha}{\rho_{\rm cl}}\ .
\end{displaymath}
The meaning of the variables is given in Table~\ref{tab:parameters}.
In the case of a selfgravitating gas disk ($\Sigma \gg \Sigma_{*}$) or a dominating
stellar disk ($\Sigma_{*} \gg \Sigma$) the set of equations can be solved analytically
(we use $\rho v_{\rm turb}^{2}  =  \pi G \Sigma^{2}$ and $\rho v_{\rm turb}^{2}  =  \pi G \Sigma \Sigma_{*}$).
In Appendix~\ref{sec:modelequations} we give the equations relating several observables ($v_{\rm turb}$,
$\Sigma$, $SFE = \dot{\Sigma_{*}}/\Sigma$, and $f_{\rm mol} $ and 
$f_{\rm mol}=\Sigma_{{\rm H}_2}/(\Sigma_{{\rm H}_2}+\Sigma_{{\rm HI}})$) and other 
quantities of interest ($l_{\rm driv}$, $\nu$, $\Phi_V$) to our free parameters ($\dot{M}$, $\delta$), 
assumed constants ($\xi$, $\eta$), the approximated metallicity ($\alpha$), the ``fundamental'' observables (stellar surface density
$\Sigma_{*}$, radius $R$ and rotation $\Omega$), and the ``quasi-free'' parameter Toomre $Q$ of the gas (which we treat as free, but
compare to observations).

For the global comparison between the observed and the model radial profiles we solve the set of equations
given above numerically and we use 
\begin{equation}
\frac{1}{t_{\rm SF}^{\rm l}}=\frac{1}{\sqrt{\frac{3\pi}{32G\rho_{\rm cl}}+\alpha^{2}/\rho_{\rm cl}^{2}}}\ .
\end{equation}

\section{Comparison with Observations}

We compare the model described above to radial profiles of gas and
star formation for 18 nearby spiral galaxies.

\subsection{Data \label{sec:data}}

The galaxies we compare to are all part of the THINGS
\citep{Walter} survey. To carry out the comparison, we use radial
profiles of the H{\sc i} velocity dispersion measured via the second
moment presented by \citet{Tamburro}; rotation curves derived by
\citet{deBlok}; radial profiles of atomic (H{\sc i}) and molecular gas
(traced via CO) surface density based on maps by \citet{Walter},
\citet{Leroy09}, and \citet{Helfer03}; and profiles of star formation
rate surface density derived by \citet{Leroy} from a combination of
24$\mu$m \citep[SINGS,][]{Kennicutt03} and FUV \citep[the GALEX
NGA,][]{GilDePaz} intensity. We use stellar surface density estimates
derived from SINGS $3.6\mu$m imaging by \citet{Leroy} assuming a
constant mass-to-light ratio.

The profiles of atomic and molecular gas are combined into profiles of
the molecular fraction, $f_{\rm mol}=\Sigma_{\rm H_2}/(\Sigma_{\rm
H_2}+\Sigma_{\rm HI})$, and the profiles of star formation rate, H{\sc
i}, and molecular gas are combined into profiles of star formation
efficiency, $SFE = \dot{\Sigma}_{\rm *} / \Sigma$. Most of the profiles
are drawn from \citet{Leroy} and details on methodology and data can
be found there.

The following uncertainties affect the data in addition to the
uncertainty in the mean value over the azimuthal ring estimated from
the RMS scatter in that ring:

{\em SFR and SFE}: there is a $\sim 50\%$ uncertainty based on
inter-comparison of different star formation rate tracers. Half of this
($\sim 0.1$~dex) is internal to the galaxy, the other half ($\sim
0.1$~dex) is galaxy-to-galaxy scatter (due to, e.g., the star
formation history and dust properties).

{\em \sc Hi}: THINGS is expected to recover the true {\sc Hi} flux
with 10\% accuracy.

{\em CO}: The global CO-H$_2$ conversion factor is uncertain by a
factor of 2, which dwarfs any uncertainty in the calibration.  An
additional systematic uncertainty is the dependence of the conversion
factor on local conditions. The sample selection somewhat minimizes
this concern. The CO data have been masked before constructing
profiles, which can lead to artificially steep profiles at the edge of
the CO emitting region.

{\em Velocity dispersion}: the scatter in the dispersion within a
given ring is about 30\%. Five of the 18 galaxies are too inclined to
derive reliable velocity dispersions. Several regions have very high
velocity dispersions, $>20$~km~s$^{-1}$, which we expect are due at
least partially to non-circular motions like bars, outflows, or
streaming along spiral arms. We do not use these data in the
comparison.

\subsection{Procedure}

The comparison is done in the following way:
\begin{enumerate}
\item
The Toomre parameter of the gas is calculated using the observed H{\sc i} dispersion velocity $v_{\rm disp}$,
rotation velocity $v_{\rm rot}$, and total gas surface density $\Sigma_{\rm gas}$:
\begin{equation}
Q=\frac{v_{\rm disp} v_{\rm rot}}{\pi G \Sigma_{\rm gas} R}\ .
\end{equation}
This approximate form of $Q$ is used to avoid taking
the (noisy) derivative of the observed rotation curve. In two
cases, we adjust $Q$ somewhat from the measured value: 
(i) in the inner part ($R \leq 4$~kpc) of the disk of NGC~3351 where we set $Q=8$
and (ii) in the whole disk of NGC~2841 where we also assume $Q=8$.
For both galaxies this corresponds to an approximately constant
velocity dispersion of $\sim 12$-$15$~km\,s$^{-1}$.
In cases without a measured velocity dispersion, we adopt a flat $Q$ (NGC~2841, NGC~3351), 
an exponentially declining $Q$ (NGC~3627), or a constant velocity dispersion of 15~km\,s$^{-1}$ (NGC~3521, NGC~7331).
\item
We solve the set of equations given in Sect.~\ref{sec:calc} using a grid for values of $\delta$,
$\dot{M}$, and $\gamma$. To account for the overall calibration uncertainties (Sect.~\ref{sec:data}), we calculate the
models on the $\delta$-$\dot{M}$-$\gamma$-grid multiplying all points of $\Sigma_{\rm H_2}$ by $(0.63,1.,1.58)$,
$v_{\rm turb}$ by $(0.79,1.,1.26)$, and $SFE$ by $(0.71,1.,1.41)$, allowing for all permutations.
For each set of parameters the goodness $g$ of the fit is determined for each observable profile:
\begin{equation}
g = \frac{1}{N} \sum_{i=1}^{N} \frac{(V^{\rm obs}_{i}-V^{\rm model}_{i})^2}{\sigma^2}\ ,
\end{equation}
where $N$ is the number of points of a radial profile, $V_{\rm obs/model}$ are the observed and
the model points of the profiles, and $\sigma$ is the uncertainty of a point given by \citet{Leroy}.
At the end we calculate a total goodness of the fit by taking the mean of the goodnesses of the individual 
profiles. This does not correspond to a traditional $\chi^2$, because the uncertainties of the points of a given
profile are correlated.
\item
The final $\delta$ and $\dot{M}$ are determined by the mean of all $\delta$ and $\dot{M}$ yielding with
$\min(g) \leq g \leq 1.1\,\min(g)$. We verified that these values are not 
significantly different from those derived from the minimum of the total goodness.
In practice, the velocity dispersion 
provides the strongest constraint on $\dot{M}$, because it has a
small error (it is shown in linear scale in Figs.~\ref{fig:radialprofilesKM05} and
\ref{fig:radialprofilesVB03}), followed by the
molecular fraction and the SFE. Increasing $\dot{M}$ leads to an
increasing velocity dispersion (Eq.~\ref{eq:vturbsgz3} and
\ref{eq:vturbsgz}) and decreasing $f_{\rm mol}$ (Eq.~\ref{eq:fmol3}
and \ref{eq:fmol}) and SFE (Eq.~\ref{eq:sfe3} and
\ref{eq:sfe}). Typically, a variation in $\dot{M}$ by a factor of 3
leads to changes in $v_{\rm turb}$, $f_{\rm mol}$, and SFE that are
larger than the error bars (cf. Sect.~\ref{sec:how}).
\end{enumerate}

\section{Results \label{sec:results}} 

We have applied the VB03 model using the KM05 (Eq.~\ref{eq:starform}) and VB03 (Eq.~\ref{eq:starformm})
star formation descriptions to 18 spiral galaxies of the sample of \citet{Leroy} (Table~\ref{tab:galaxies}).
The H{\sc i} rotation curves are from \citet{deBlok}.
We realized that in the framework of the VB03 star formation prescription the free fall times of the largest 
selfgravitating clouds are significantly larger than the collapse times derived by \citet{Tamburro08} 
for NGC~0925, NGC~2403, and NGC~7793.
Since the model fits are sensitive to the local vertical pressure equilibrium and the assumed
correlation between the vertical and radial stellar scale height has a large scatter,
we prefer a gravitational potential dominated by the stellar disk for these galaxies yielding
local free fall times in agreement with \citet{Tamburro08}.

The resulting radial profiles are presented in Fig.~\ref{fig:radialprofilesKM05} for the KM05 star formation prescription
and in Fig.~\ref{fig:radialprofilesVB03} for the VB03 star formation prescription. The order of the plots 
(from top left to bottom right) is: H{\sc i} rotation curve, observed and fitted stellar surface density profile $\Sigma_{*}$
(the assumed $\Sigma_{*}$ for the fit is shown as a solid line),
model (dashed) and observed (solid) total gas density profile $\Sigma$, Toomre $Q$ parameter of stars+gas
(dash-dotted) and of the gas
derived from observations (dotted) and assumed for the fit (solid),
driving scale length $l_{\rm driv}$, H{\sc i} velocity dispersion (solid) and model turbulent velocity dispersion (dashed),
observed (solid) and modeled (dashed) star formation efficiency, observed (solid) and modeled (dashed) molecular fraction,
star formation (solid) and viscous (dashed) timescales, free fall timescale of the most massive
selfgravitating gas clouds, radial velocity of the gas within the disk, and the observed (solid) and
model (dashed) metallicity. The observed metallicity profiles are derived from $12+\log({\rm O/H})$
\citep{Moustakas} assuming a solar oxygen abundance of $12+\log({\rm O/H})=8.9$.
For the calculation of the combined Toomre parameter (stars+gas) $Q_{\rm tot}$ \citep{Rafikov} we limited the scale of 
instabilities to $3$~kpc.

In agreement with \citet{Leroy} we find that $Q_{\rm tot}$ is close to one for most of the galaxies. 
Only NGC~3521 and NGC~7331 show $Q_{\rm tot} \sim 2$ and for NGC~2841 $Q_{\rm tot} \sim 3$. 
The Toomre parameter of the gas ranges between 2 and 4 for most of the galaxies and these galaxies
can thus be assumed in a quasi equilibrium state 
where the mass accretion (radially in the disk and external) balances star formation to keep $Q_{\rm tot} \sim 1$.
Moreover, for all galaxies with masses smaller than $10^{10}$~M$_{\odot}$, the viscous timescale is
smaller than or comparable to the star formation timescale. For these galaxies our assumption of
a stationary accretion disk (Eq.~\ref{eq:transport}) is justified without the need of a high external accretion rate. 
For more massive galaxies Eq.~\ref{eq:transport} is a convenient parametrization which leads to a meaningful mean 
viscosity as discussed in Sect.~\ref{sec:model}.
The parameters derived from the fitting procedure are the disk mass accretion rate $\dot{M}$, the scaling between
the driving and dissipation length scale $\delta$, 
and the break radius of the star formation efficiency $R_{\rm break}$ which are given in Table~\ref{tab:galaxies1} for the
KM05 and VB03 star formation prescriptions.

We find acceptable fits for all galaxies (KM05 and VB03), except for NGC~5194 (M~51).
NGC~2976, NGC~4736, NGC~5194, NGC~3521, and NGC~5055 show $g > 1.5$.
For NGC~4736 we suspect a different mass accretion rate between the inner ($R < 2$~kpc)
and outer disk. This would violate our assumption of stationarity, i.e. a radially
constant mass accretion rate.
For NGC~5055 ($g \sim 4$), the deviations between the model and the observations occur in the
central part of the galaxies where the velocity dispersion is overestimated. For NGC~2976 and NGC~3521 the
disagreement between the model and observations occurs in the outer disk region.
The KM05 star formation prescription provides better fits (smaller $g$) for 4 galaxies 
(NGC~4214, NGC~4736, NGC~3351, and NGC~5194).
The profiles of the other 14 galaxies are better fitted by the VB03 star formation prescription.

Whereas the KM05 star formation prescription yields a constant $\delta=1.9 \pm 0.5$, $\delta$ increases 
with galaxy mass in the VB03 star formation prescription. 
Introducing a $\gamma > 0$ leads to a significant improvement of the fit (about a two times smaller goodness).
We find $\gamma=0.10 \pm 0.03$ for the KM05 star formation prescription. VB03 shows a bimodal 
distribution of $\gamma$
with 5 galaxies having $\gamma=0.33$ and 6 galaxies having $\gamma=1.0$. 
Overall $\gamma=0.65 \pm 0.34$ for the VB03 star formation prescription.
This translates into $R_{\rm break}/l_{*}=3.0 \pm 1.1$ and $R_{\rm break}/R_{25}=0.7 \pm 0.2$
for the KM05 star formation prescription and for massive galaxies ($\log(M_{*})>10$) $R_{\rm break}/l_{*}=2.0 \pm 1.0$ and 
$R_{\rm break}/R_{25}=0.4 \pm 0.2$ for the VB03 star formation prescription.
The break radius of the VB03  star formation prescription is thus close to the transition 
between a mostly-H{\sc i} and a mostly-H$_2$ ISM
\citep{Leroy}, whereas it is closer to the optical radius in the KM05 star formation prescription.

{\em Total gas surface density:} For most of the sample galaxies the total gas surface density profile is well reproduced.
In the framework of the KM05 star formation prescription the total gas surface density is underestimated
in the outer disk of IC~2574, NGC~7793, and NGC~0925; it is overestimated in the outer disk
of NGC~4736. In the framework of the VB03 star formation prescription the total gas surface density is somewhat underestimated
in IC~2574, NGC~0925, and greatly underestimated in NGC~5194; it is overestimated in the outer disk of NGC~4736.
All these differences are within $\sim 30\%$, except for NGC~5194 for the VB03 star formation prescription.

{\em Velocity dispersion:} Velocity dispersion profiles are available
for 13 out of our 18 sample galaxies.  We do not try to fit dispersion
velocities higher than 20~km\,s$^{-1}$. These tend to appear in
  regions where geometry confuses the measurement (e.g., the central
  parts of inclined disks) or regions of very high SFR, where we worry
  that outflows or other non-disk structures render the measured
  velocity dispersion inappropriate for comparison to the model. For
  examples of our concerns, see the position-velocity diagrams of
  \citet{deBlok} or compare the results of gaussian fitting to moment
  methods treating the same galaxy
  \citep[e.g.,][]{Boomsma,Tamburro}. Both reveal the presence of
  asymmetric, multi-component {\sc Hi} profiles in the central parts
  of some galaxies. We can reproduce the observed profiles within the
error bars for all galaxies (KM05 and VB03 star formation
prescriptions), except NGC~5194 with the VB03 star formation
prescription.

{\em Star formation efficiency:} The star formation efficiency in the inner disk of all galaxies is
well reproduced by both models. The star formation efficiency of the outer disks of NGC~2976, NGC~0925,
NGC~3198, NGC~5055, and NGC~7331 is better reproduced by the VB03  than by the KM05 star formation prescription. 
On the other hand,
that of NGC~4214, NGC~6946, NGC~5194, and NGC~2841 is better reproduced by the KM05 star formation prescription.
NGC~2976 and NGC~4736 show very low surface density gas in this region leading to $Q > 10$. 
Since the star formation efficiency is proportional to $Q^{-1}$ in the outer disk, the mismatch
between model and data might be due to our incomplete knowledge of $Q$ in these regions.

{\em Molecular fraction:} For 15 out of the 18 sample galaxies CO observations are available. 
The VB03 star formation prescription reproduces
the molecular fractions of almost all galaxies (except NGC~4736 and NGC~5194) roughly within the error bars.
There are deviations between the KM05 star formation prescription and observed molecular fractions for NGC~5194, NGC3521, NGC~5055.
In NGC~3184, NGC~4736, NGC~6946, and NGC~5194 the
decline of the observed molecular fraction at the outer edge of the CO distribution is much steeper 
than predicted by the model. This is mostly due to sensitivity, because the CO data cubes were clipped at
given S/N levels.

\section{Discussion \label{sec:discussion}}

Since the VB03 model is axisymmetric, it does not include non-axisymmetric structures like
spiral arms and bars. Whereas spiral arms do not change the given picture
considerably \citep{Haan}, the existence of bars do have a strong impact on radial gas flows.
Despite the importance of a bar in the evolution of galactic disks,
the fact that the VB03 model does not include their effect does not affect our conclusions.

In the following we discuss aspects related to different radial profiles:

{\em Star formation law:}
Fig.~\ref{fig:hibigiel} shows the star formation rate per unit area as a function of the
total gas surface density (H{\sc i}+H$_2$) of our model. The distribution agrees well
with observed distributions from THINGS and the literature \citep[Fig.~15 of][]{Bigiel}.
The absolute values and the shape, i.e. the 'knee', are reproduced by both the KM05 and VB03 models.

{\em Velocity dispersion:}
As shown in \citet{Tamburro}, the velocity dispersion of spiral galaxies decreases with increasing 
galactocentric radius. In the following, 
We note that the VB03 model for a stellar disk dominated gravitational potential yields a radially
declining velocity dispersion, whereas a selfgravitating gas disk leads to a constant velocity dispersion 
in the inner disk  ($R < R_{\rm break}$) and a declining velocity dispersion in the outer disk ($R > R_{\rm break}$).
Thus, it is in principle possible to determine these different regimes from the
radial behavior of the observed velocity dispersion.

{\em Molecular fraction:} With the model radial profiles we can
investigate the dependence of the star formation rate per unit surface
area on the molecular and total surface density
(Fig.~\ref{fig:molbigiel}).  The correlation between the star
formation rate and the molecular gas surface density is linear with a
molecular depletion timescale of $\sim 1.5$~Gyr for the KM05 star
formation prescription and $\sim 2$~Gyr for the VB03 star formation
prescription.  The correlation between the total gas surface density
and the star formation rate is steep where H{\sc i} is dominating
($\Sigma < 20$~M$_{\odot}$yr$^{-1}$) and flattens for higher gas
surface densities.  The VB03 model reproduces results by
\citet{Bigiel} and \citet{Leroy} who found (i) an average molecular
gas depletion timescale of $2$~Gyr and (ii) a critical gas surface
density of $14$~M$_{\odot}$pc$^{-2}$ for the change between gas
predominantly in molecular form and gas predominantly in atomic form.

{\em Driving length scale:}
The derived driving scale lengths are monotonically increasing with increasing radius.
Typical values are $100$-$300$~pc in the inner half of the optical disk. 
At the optical radius the driving length scale is about $400$-$800$~pc.
The radial increase and the large values at the optical radius are consistent with sizes of {\sc Hi}
shells observed in the Galaxy \citep{Heiles,McClure}.

\begin{figure*}
\epsscale{.8}
\plotone{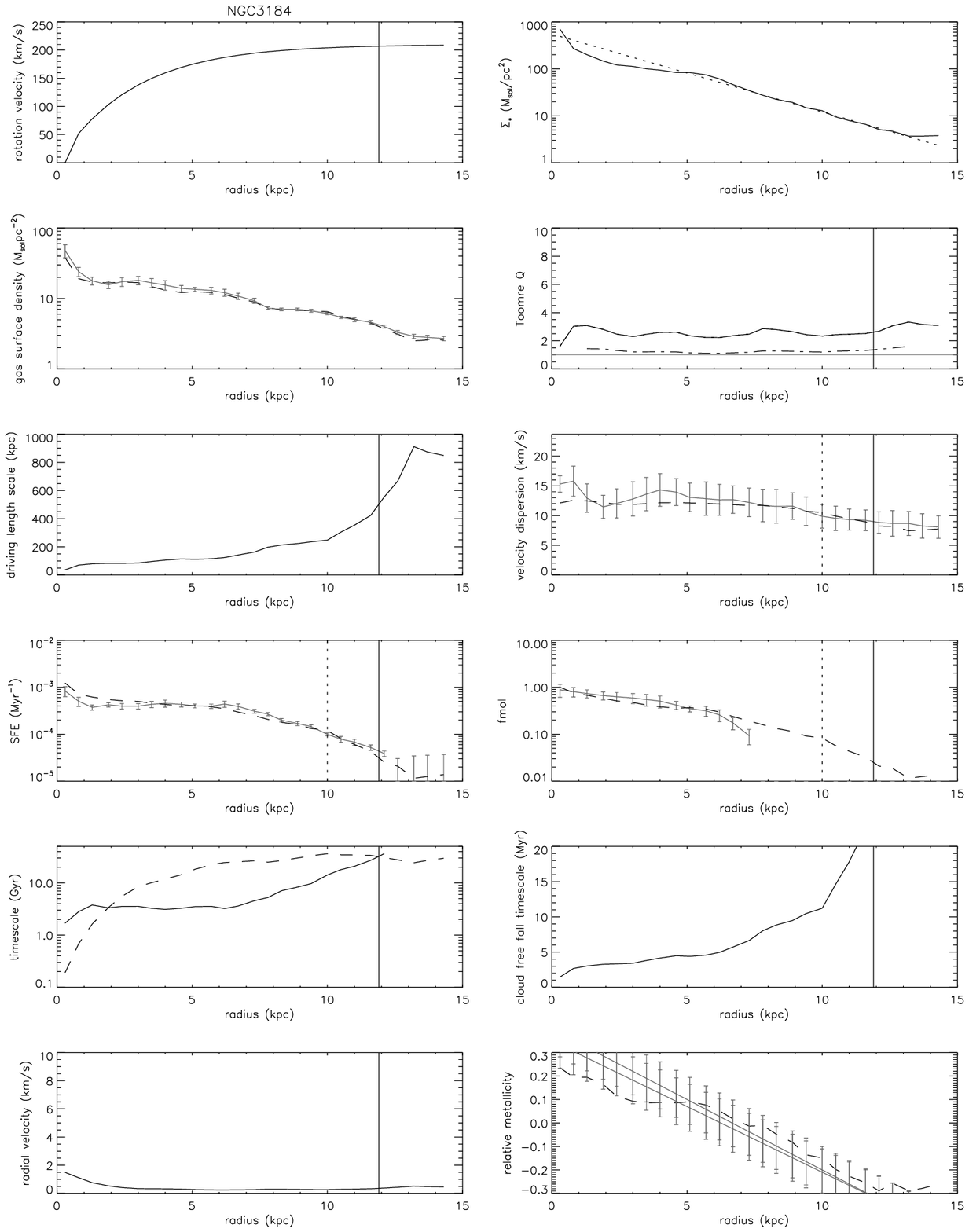}
\caption{Comparison between modeled (KM05) and observed radial profiles. Form upper left to
lower right: (i) H{\sc i} rotation curve, (ii) observed and fitted stellar surface density profile $\Sigma_{*}$
(the assumed $\Sigma_{*}$ for the fit is shown as a solid line),
(iii) model (dashed) and observed (solid) total gas density profile $\Sigma$, (iv) Toomre $Q$ parameter of stars+gas
(dash-dotted) and of the gas
derived from observations (dotted) and assumed for the fit (solid),
(v) driving scale length $l_{\rm driv}$, (vi) H{\sc i} velocity dispersion (solid) and model turbulent velocity dispersion (dashed),
(vii) observed (solid) and modeled (dashed) star formation efficiency, (vii) observed (solid) and modeled (dashed) molecular fraction,
(ix) star formation (solid) and viscous (dashed) timescales, (x) free fall timescale of the most massive
selfgravitating gas clouds, (xi) radial velocity of the gas within the disk, and (xii) the observed (solid) and
model (dashed) metallicity. The observed metallicity profiles is derived from $12+\log({\rm O/H})$
\citep{Moustakas} assuming a solar oxygen abundance of $12+\log({\rm O/H})=8.9$. The two solid lines
correspond to the calibrations of \citet{Kobulnicky} and \citet{Pilyugin} with an additional offset of 0.6~dex.
\label{fig:radialprofilesKM05}}
\end{figure*}

\begin{figure*}
\epsscale{.8}
\plotone{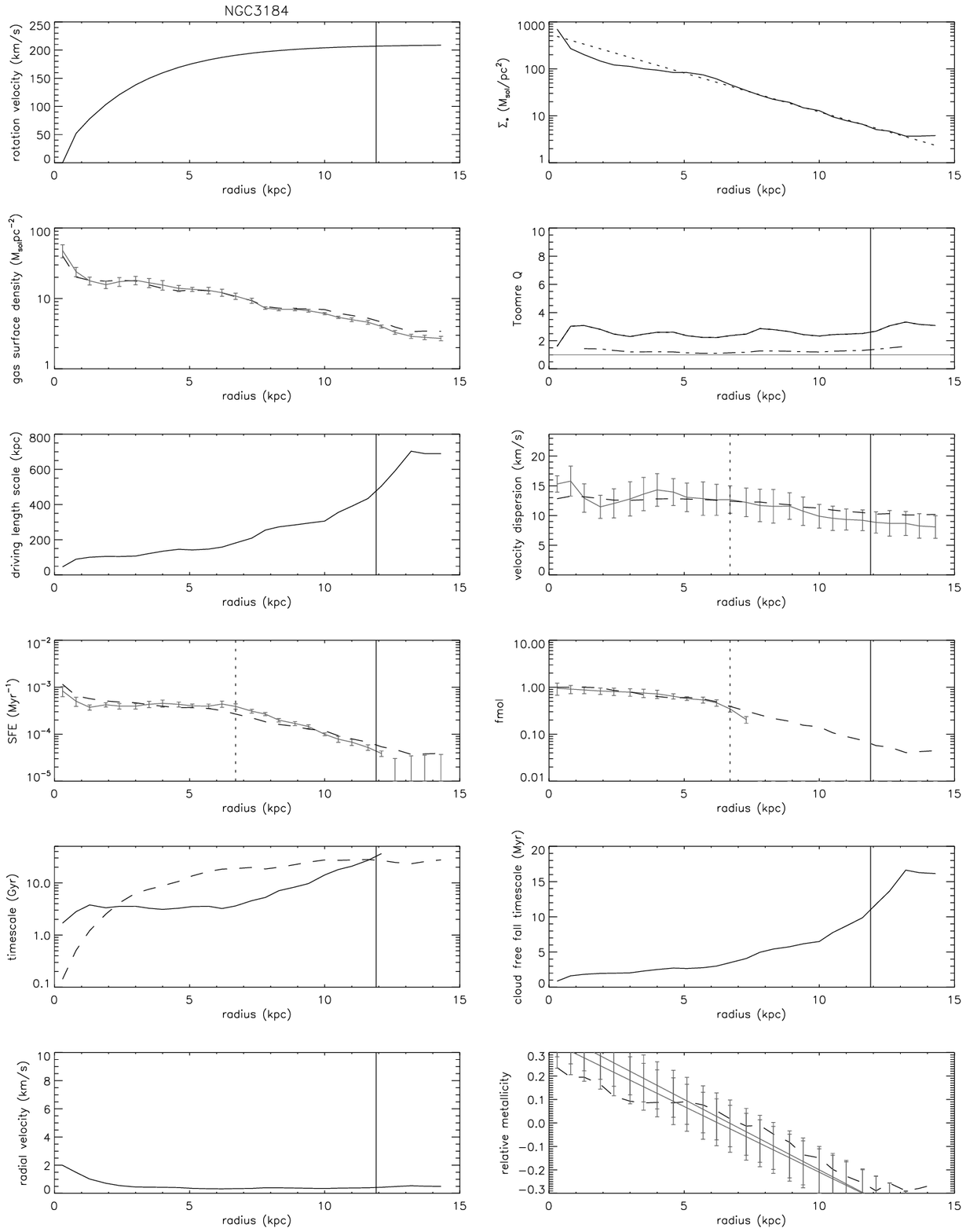}
\caption{Comparison between modeled (VB03) and observed radial profiles. Form upper left to
lower right: (i) H{\sc i} rotation curve, (ii) observed and fitted stellar surface density profile $\Sigma_{*}$
(the assumed $\Sigma_{*}$ for the fit is shown as a solid line),
(iii) model (dashed) and observed (solid) total gas density profile $\Sigma$, (iv) Toomre $Q$ parameter of stars+gas
(dash-dotted) and of the gas
derived from observations (dotted) and assumed for the fit (solid),
(v) driving scale length $l_{\rm driv}$, (vi) H{\sc i} velocity dispersion (solid) and model turbulent velocity dispersion (dashed),
(vii) observed (solid) and modeled (dashed) star formation efficiency, (vii) observed (solid) and modeled (dashed) molecular fraction,
(ix) star formation (solid) and viscous (dashed) timescales, (x) free fall timescale of the most massive
selfgravitating gas clouds, (xi) radial velocity of the gas within the disk, and (xii) the observed (solid) and
model (dashed) metallicity. The observed metallicity profiles is derived from $12+\log({\rm O/H})$
\citep{Moustakas} assuming a solar oxygen abundance of $12+\log({\rm O/H})=8.9$. The two solid lines
correspond to the calibrations of \citet{Kobulnicky} and \citet{Pilyugin} with an additional offset of 0.6~dex.
\label{fig:radialprofilesVB03}}
\end{figure*}

\begin{figure*}
\epsscale{.9}
\plottwo{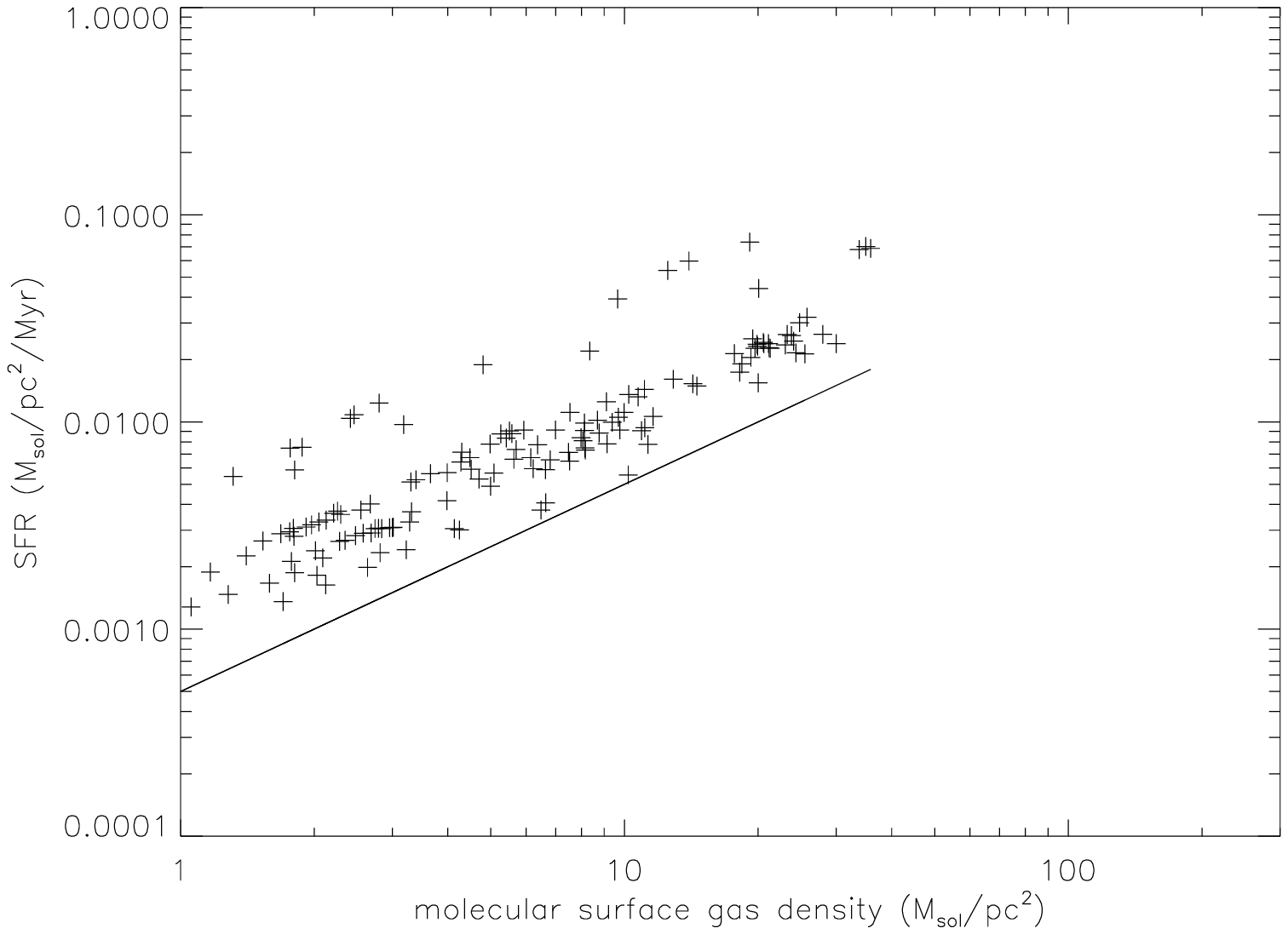}{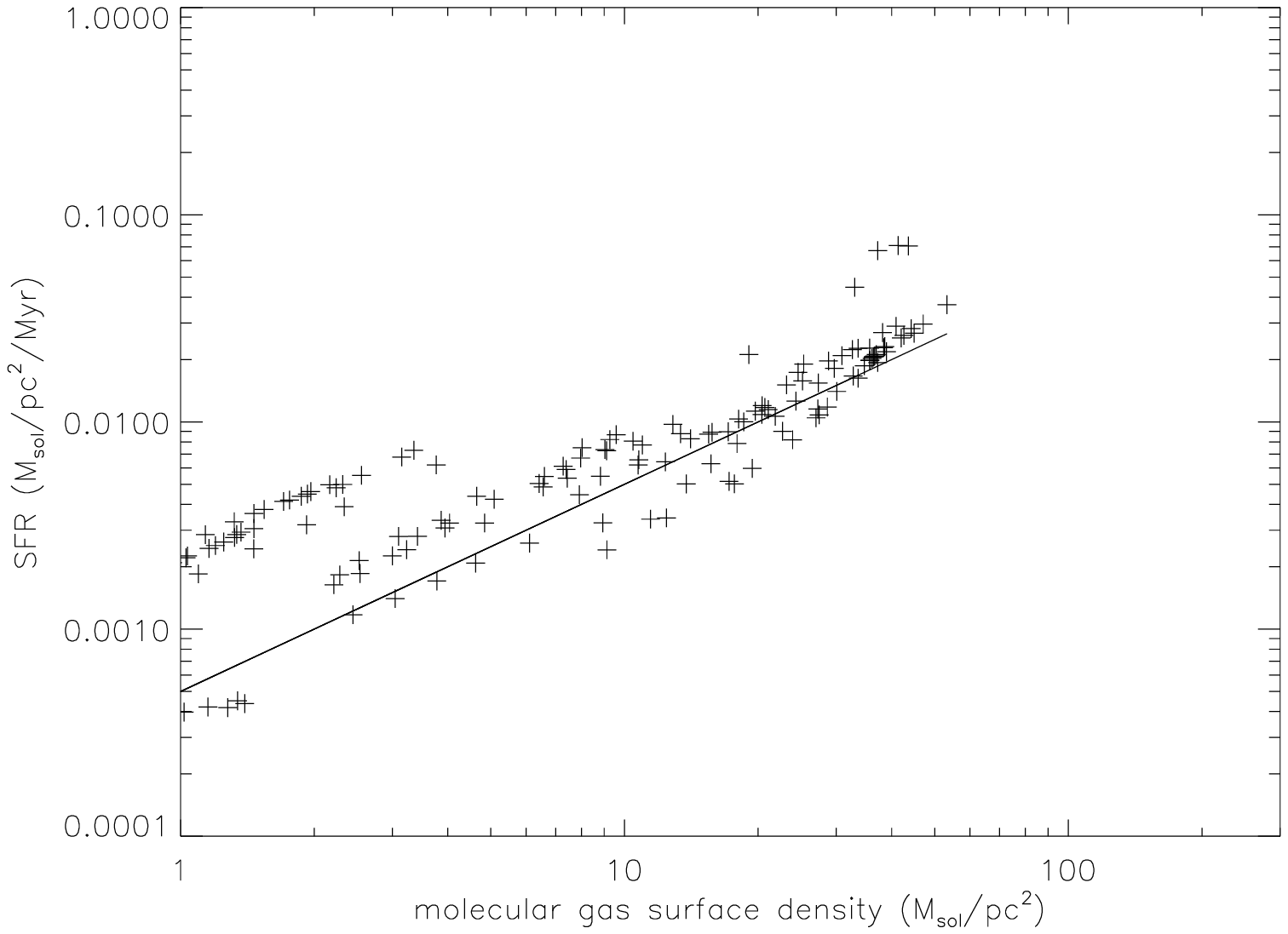}
\caption{Correlation between the star formation rate per unit surface area and
the molecular gas surface density.  
The solid line correspond to a molecular gas depletion rate of $2$~Gyr.
Left panel: KM05 star formation prescription. Right panel: VB03 star formation prescription.
\label{fig:molbigiel}}
\end{figure*}

\begin{figure*}
\epsscale{.9}
\plottwo{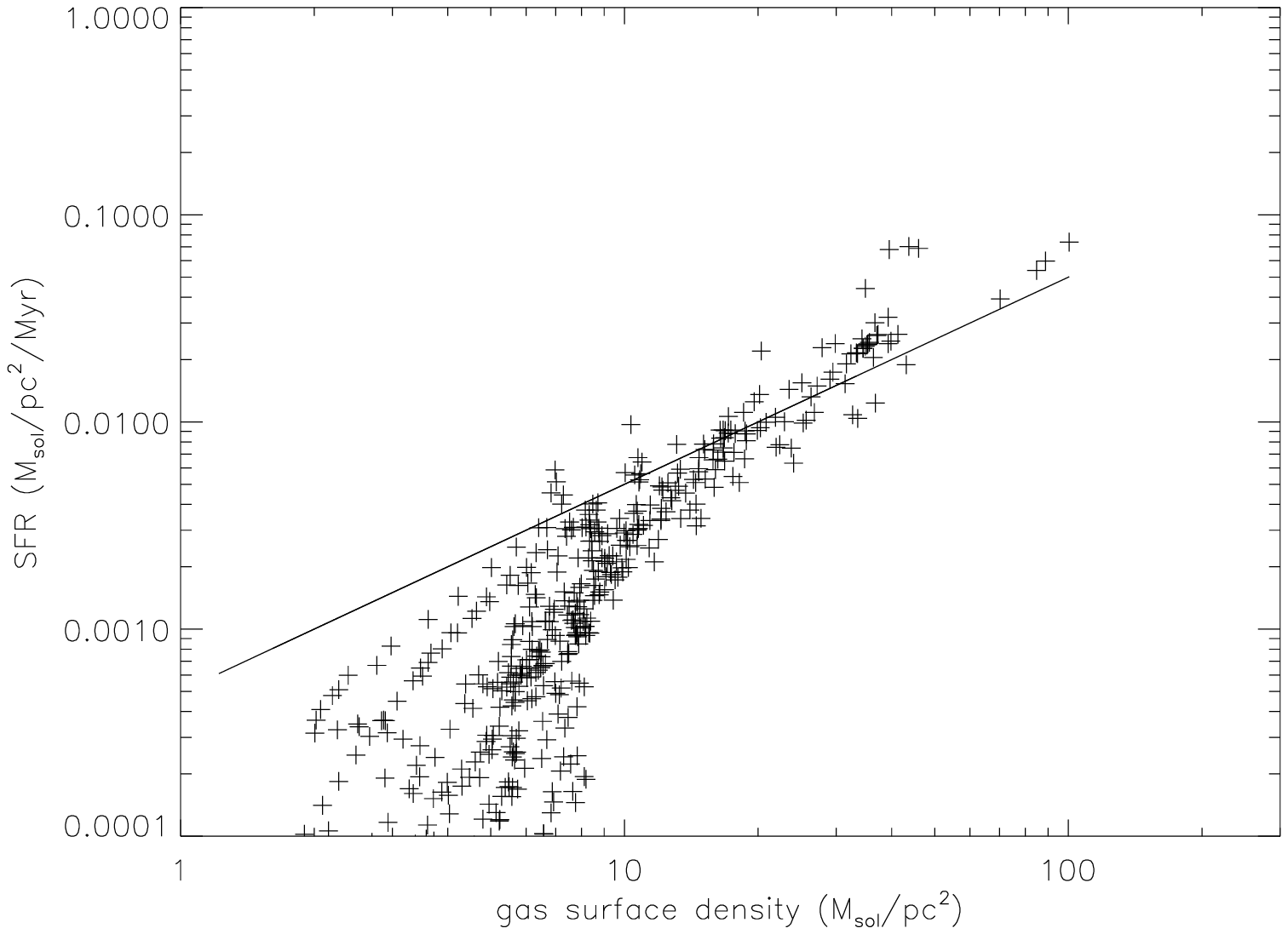}{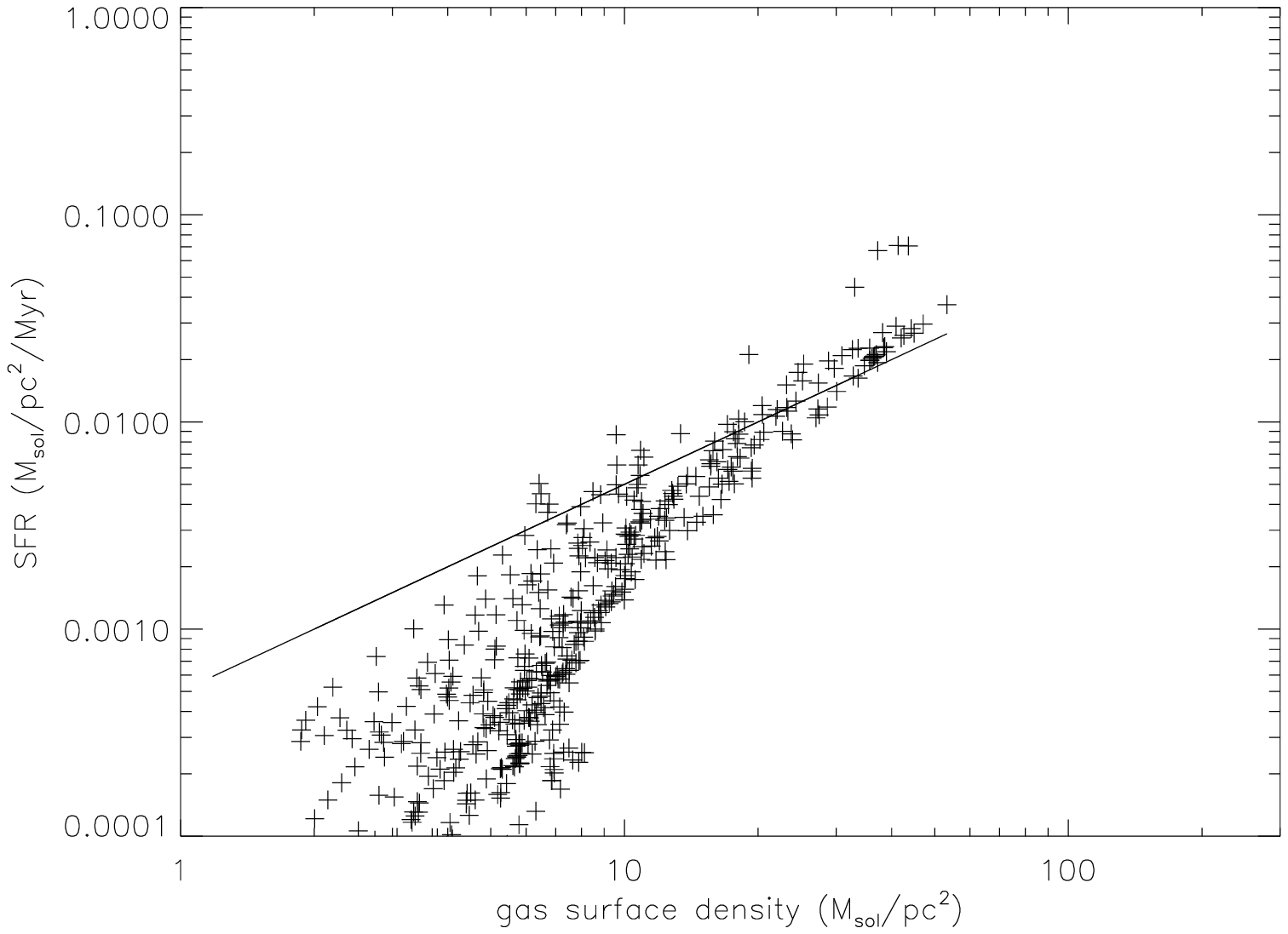}
\caption{Correlation between the star formation rate per unit surface area and
the total gas surface density.  
The solid line correspond to a molecular gas depletion rate of $2$~Gyr.
Left panel: KM05 star formation prescription. Right panel: VB03 star formation prescription.
\label{fig:hibigiel}}
\end{figure*}

{\em Radial gas motions:}
Radial gas motions with velocity $v_{\rm rad}$ can be estimated with the mass accretion rate $\dot{M}$ and 
the gas surface density $\Sigma_{\rm gas}$ \citep[see e.g.][]{Pringle}:
\begin{equation}
\dot{M}=2 \pi R \Sigma_{\rm gas} (-v_{\rm rad})
\end{equation}
The radial profiles of the radial velocity are presented in Figs.~\ref{fig:radialprofilesKM05}
and \ref{fig:radialprofilesVB03}. Typically, we find radial 
velocities smaller than $2$~km\,s$^{-1}$. This is consistent with the findings of
\citet{Trachternach} who found non-circular motions smaller than $5$~km\,s$^{-1}$ in these galaxies.
Moreover, we observe a general increase of radial velocities with decreasing galactocentric radius.
A strongly increasing $Q$ in the outer disks of NGC~2976, NGC~4736, and NGC~3627 due to a
strongly decreasing gas surface density leads to steeply increasing radial velocities
(up to $10$~km\,s$^{-1}$) in these galaxies. This is again consistent with the profiles
of non-circular motions derived by \citet{Trachternach}.

{\em Cloud free fall time estimates:}
The free-fall time only depends on the cloud density $\rho_{\rm cl}$ (Eq.~\ref{eq:localff}). 
Typical densities of giant molecular clouds are between $100$-$1000$~cm$^{-3}$ 
\citep{Solomon, Heyer} leading to free fall times of $1.6$-$5.1$~Myr.
\citet{Tamburro08} estimated a characteristic timescale for star formation in the spiral arms of 
disk galaxies, going from atomic hydrogen (H{\sc i}) to dust-enshrouded massive stars.
Their free fall time estimates vary between $1$ and $6$~Gyr.
The VB03 model yields the physical properties of the most massive selfgravitating clouds at a given
galactocentric radius via the local free-fall ($t_{\rm ff}^{l}$), turbulent ($t_{\rm turb}^{l}$), and
molecule formation ($t_{\rm mol}^{l}$) timescales. We recall that for selfgravitating clouds
$t_{\rm ff}^{l}=t_{\rm turb}^{l}$. The radial profiles of the local free fall time for our sample
galaxies are shown in Figs.~\ref{fig:radialprofilesKM05} and \ref{fig:radialprofilesVB03}. 
They are calculated using Eq.~\ref{eq:localff} and \ref{eq:phiv}.
For the comparison with the free fall timescales derived from observations we calculated
the mean and the standard deviation of the radial profiles within $l_* \leq R \leq 2\,l_*$.
Our free fall timescales (Table~\ref{tab:galaxies1}) 
are in good agreement with those derived from observations except for those of NGC~2403, NGC~0925,
(both star formation prescriptions) and NGC~7793 (VB03 star formation prescription). 
By modifying Eq.~\ref{eq:alphacb} to better match the observed
metallicity profile:
\begin{equation}
\alpha=1.6 \times 7.2 \times 10^{7} \times \big( \log(\frac{\Sigma_{*}+\Sigma}{\Sigma}) \big)^{-1}\ 
{\rm yr\,M_{\odot}pc^{-3}}\ ,
\end{equation}
we obtain the following parameters: NGC~7793: $\delta=1$, $\dot{M}$=0.06~M$_{\odot}$yr$^{-1}$;
NGC~2403: $\delta=4$, $\dot{M}$=0.22~M$_{\odot}$yr$^{-1}$; NGC~0925: $\delta=9$, $\dot{M}$=0.31~M$_{\odot}$yr$^{-1}$.
With these parameters the free fall timescales of these three galaxies are in good agreement with
expectations. We note that in this case $\gamma=0.33$ for NGC~7793 and NGC~2403.

\section{How to sustain the star formation rate \label{sec:how}}

\begin{figure*}
\epsscale{1.0}
\plottwo{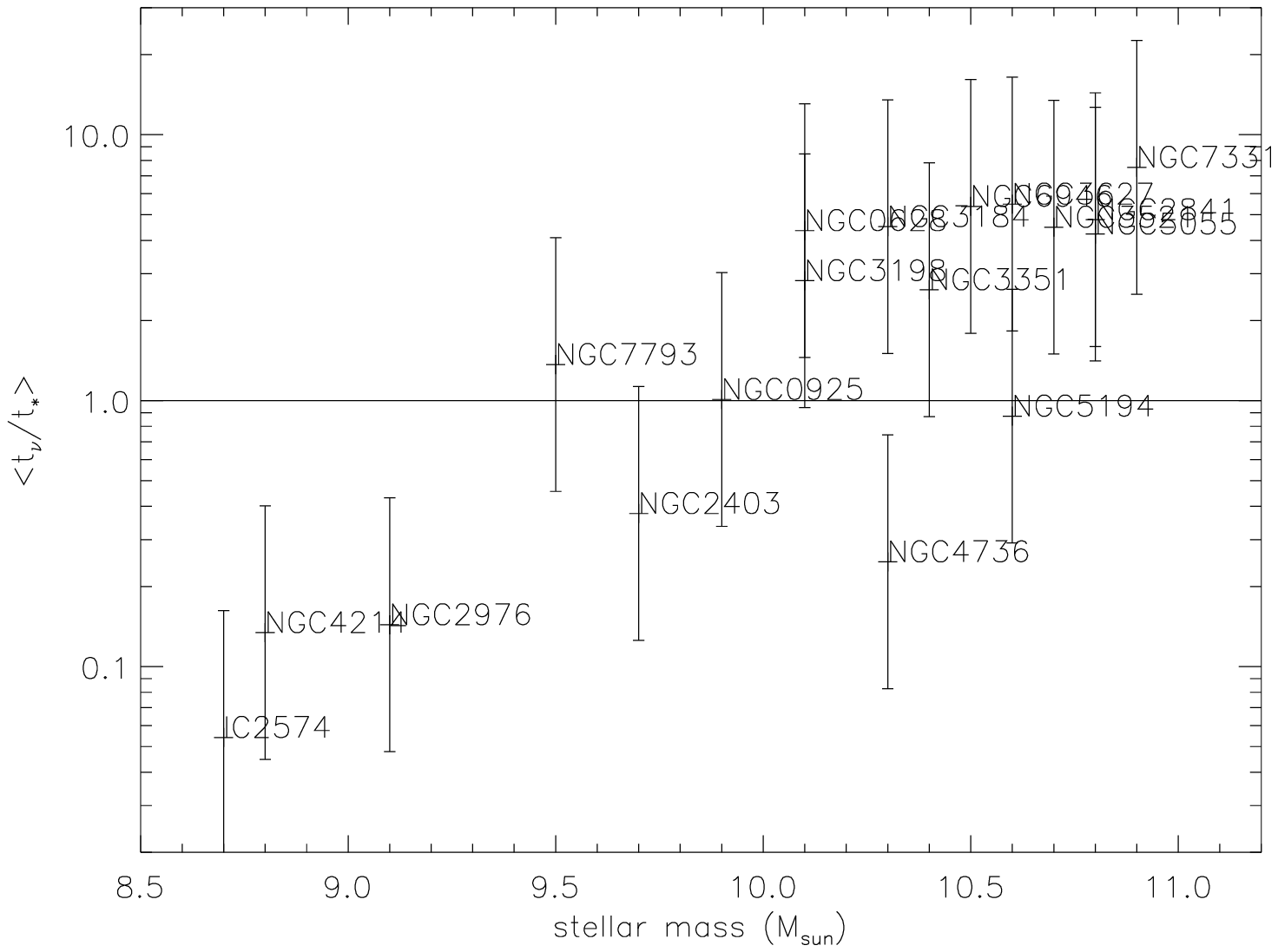}{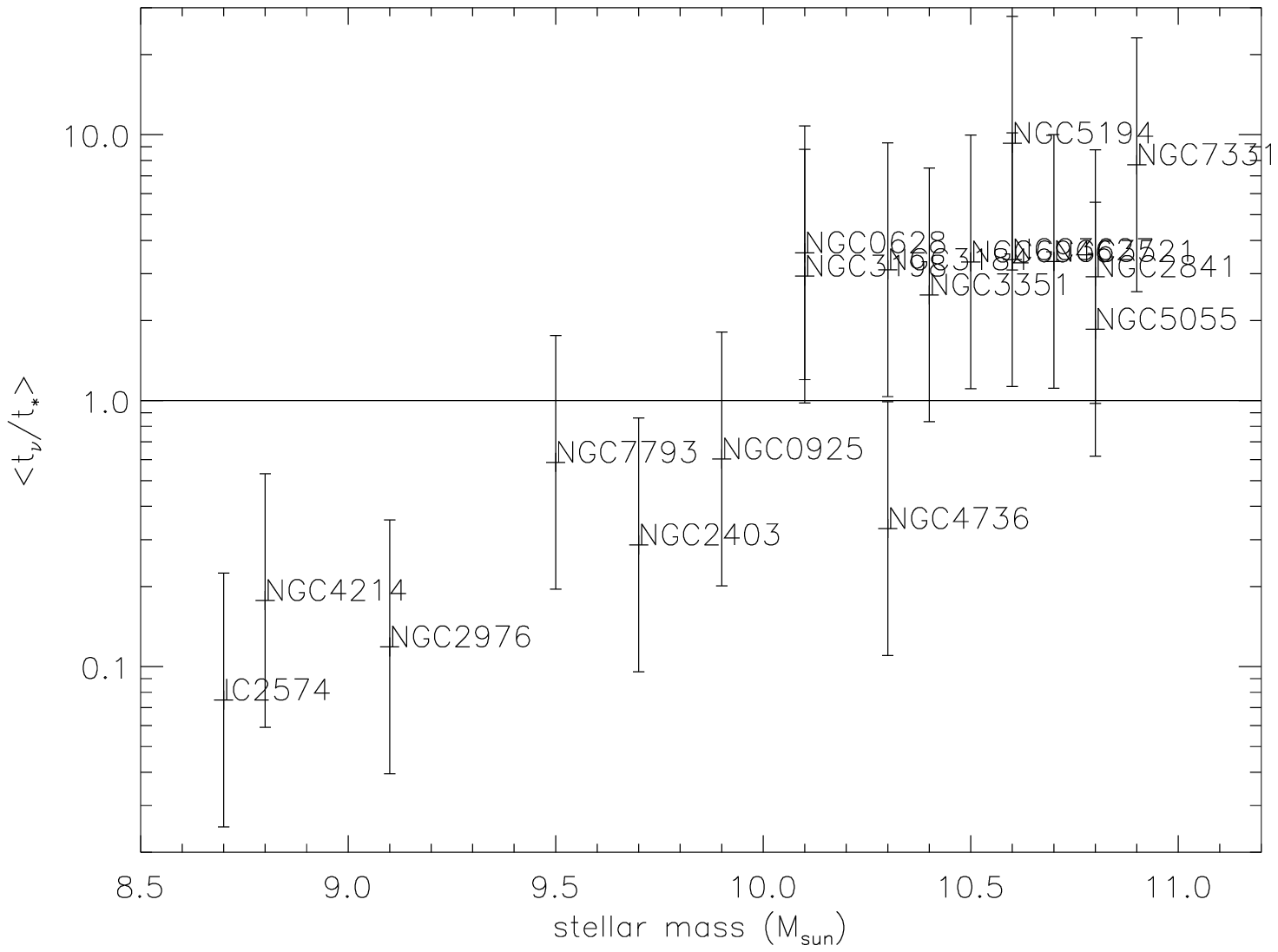}
\caption{Mean fraction between the model-derived mass accretion timescale and the observed star formation timescale
for $l_{*} \leq R \leq R_{\rm 25}$.
The error bars are based on a factor of 3 uncertainty of the mass accretion rate
(see Sect.~\ref{sec:how}). Left panel: KM05 star formation prescription. Right panel: VB03 star formation prescription. \label{fig:sfrmdot}}
\end{figure*}

Can these galaxies sustain their star formation rates by radial transport of gas within the
galactic disk? To answer this question one has to compare the local star formation rate and the local viscous
timescale $t_{\nu}=R^{2}/\nu$ using Eq.~\ref{eq:nu} and \ref{eq:nu1}.
The local timescale comparison is presented in Figs.~\ref{fig:radialprofilesKM05} and \ref{fig:radialprofilesVB03}. 
The global comparison of the mean fraction $<t_{\nu}/t_{*}>$ calculated over $l_{*} \leq R \leq R_{25}$
is shown in Fig.~\ref{fig:sfrmdot}.
The KM05 and VB03 star formation prescriptions yield consistent results for this fraction within the galactic disks.
 
The local viscous and star formation timescales are $t_{\nu}=R^2/\nu=2 \pi \Sigma R^2 / \dot{M}$ and
$t_*=\Sigma / \dot{\Sigma}_{*}$. The mean fraction is then approximately
$<t_{\nu}/t_{*}> \simeq <2 \pi \Sigma_* R^2 / \dot{M}> \simeq \dot{M}_{*}/\dot{M}$.
The important point for the global comparison is that the derived mass accretion rates for almost all
sample galaxies lie in the range between $\sim 0.1$~M$_{\odot}$yr$^{-1}$ and $0.6$~M$_{\odot}$yr$^{-1}$,
whereas the star formation rates vary between $0.1$~M$_{\odot}$yr$^{-1}$ and $3$~M$_{\odot}$yr$^{-1}$.
In the following we show why the mass accretion rate shows this behavior.
Using $\nu=v_{\rm turb}l_{\rm driv}$ in Eq.~\ref{eq:energyflux} leads to 
\begin{equation}
\label{eq:eq1}
l_{\rm driv}=\frac{\Sigma v_{\rm turb}^{3}}{\xi \dot{\Sigma_{*}}}\ .
\end{equation}
The expression for the mass accretion rate then becomes
\begin{equation}
\label{eq:eq2}
\dot{M}=2 \pi \nu \Sigma = 2 \pi v_{\rm turb} l_{\rm driv} \Sigma = \frac{2 \pi \Sigma^{2} v_{\rm turb}^{4}}{\xi \dot{\Sigma}_{*}}\ .
\end{equation}
Assuming typical values at $R_{25}/2$, $v_{\rm turb}=10$~km\,s$^{-1}$, $l_{\rm driv}=100$~pc, 
$\Sigma=10$~M$_{\odot}$pc$^{-2}$, yields a mass accretion rate of $\dot{M}=0.07$~M$_{\odot}$yr$^{-1}$.
For an error estimate we use the observational uncertainties given in Sec.~\ref{sec:data}.\\
Because of the large uncertainties associated with the molecular gas, we estimate the error
of the mass accretion rate at galactic radii larger than $R_{25}/2$ where the gas is
predominantly in atomic form. We assume
$\Sigma=10$~M$_{\odot}$pc$^{-2}$, $\Delta \Sigma=1$~M$_{\odot}$pc$^{-2}$,
$\dot{\Sigma}_{*}=5 \times 10^{-10}$~M$_{\odot}$pc$^{-2}$yr$^{-1}$, 
$\Delta \dot{\Sigma}_{*}=2.5 \times 10^{-10}$~M$_{\odot}$pc$^{-2}$yr$^{-1}$, $v_{\rm turb}=10$~km\,s$^{-1}$, and
$\Delta v_{\rm turb}=3$~km\,s$^{-1}$. 
This leads to $\log(\dot{M} ({\rm M}_{\odot}{\rm yr}^{-1}))=-0.5 \pm 0.5$, i.e. we obtain an uncertainty of about a 
factor of 3.

The inspection of Figs.~\ref{fig:radialprofilesKM05}, \ref{fig:radialprofilesVB03}, 
and \ref{fig:sfrmdot} gives the consistent answer that only the less massive
galaxies ($\log M_{*} ({\rm M}_{\odot}) \la 10$) can sustain the gas loss due to star formation by radial gas
transport within the galactic disks. These galaxies can even have access to their gas reservoir
beyond the optical radius. On the other hand, the radial gas transport in the massive spiral galaxies 
might not be sufficient to balance the gas loss due to star formation. 
This implies that, whereas a massive galaxy needs spherical infall from a putative gas halo or
has to wait for infall with an angular momentum close to that of its disk to replenish its gas content,
a less massive galaxy can live with large angular momentum accretion, because its mass accretion rate
even at $R > R_{25}$ is large enough for using this gas to sustain its star formation rate.
The star formation rate of the massive galaxies is thus set by the amount of external accretion.
In the absence of such an external gas accretion galaxies will
slowly consume their gas, the gas surface density will decrease, the Toomre parameter $Q$ of the gas will
increase, and the star formation rate will decline\footnote{This decline is somewhat slowed down
by gas replenishment from dying stars \citep{Gisler}.}. Examples in our sample might be NGC~3351 and NGC~2841. 
Given that most of the massive galaxies of our sample do not show this behavior suggests that
these galaxies experience mass accretion with rates comparable to their star formation rates (1-3~M$_{\odot}$yr$^{-1}$)
from a putative spherical halo of ionized gas or from satellite accretion leading to a
temporarily enhanced mass accretion rate within the disk.

\section{Conclusions \label{sec:conclusions}}

The theory of clumpy gas disks (VB03) provides analytic expressions
for large-scale and small-scale properties of galactic gas disks. 
Large-scale properties considered are the gas surface density, 
density, disk height, turbulent driving length scale, 
velocity dispersion, gas viscosity, volume filling factor, 
and molecular fraction. Small-scale properties are the mass, size, density,
turbulent, free-fall, molecular formation timescales of the most massive
selfgravitating gas clouds. These quantities depend on the stellar surface density,
the angular velocity $\Omega$, the disk radius $R$, and 4 free parameters, 
which are the Toomre parameter $Q$ of the gas, the mass accretion rate $\dot{M}$, 
the ratio $\delta$ between the driving length scale of turbulence and 
the cloud size, and the radius, at which the local star formation timescale
is no longer the cloud free-fall timescale, but the molecule formation
timescale.  We determine 
these free parameters using three independent measurements of the radial 
profiles of the (i) neutral gas (H{\sc i}), molecular gas (CO), and 
star formation rate (FUV + 24~$\mu$m). A sample of 18 mostly spiral galaxies 
from \citet{Leroy} is used in the analysis. Based on the simultaneous VB03 model fitting 
of the radial profiles of the total gas surface density, velocity dispersion, star formation efficiency,
and molecular fraction, we conclude that
\begin{enumerate}
\item
the model star formation efficiency is very sensitive to the description of local pressure equilibrium in the
disk midplane (Eq.~\ref{eq:pressure});
\item
the fits of all radial profiles are acceptable for all galaxies except NGC~5194 (M~51). The model-derived 
free-fall timescales of selfgravitating clouds are in good agreement with expectations from observations;
\item
the observed radially decreasing gas velocity dispersion \citep{Tamburro} can be reproduced 
by the model. In the framework of the VB03 model the decrease in the inner disk is due to the 
stellar mass distribution which dominates the gravitational potential.
A selfgravitating gas disk yields a constant velocity dispersion in the inner disk, whereas
it leads to a radially decreasing velocity dispersion in the outer disk.
It might be thus possible to identify the different regimes from the radial behavior
of the gas velocity dispersion;
\item Introducing a change in star formation regime into the model
  improves the fits significantly.  This change is realized by
  replacing the free fall time in the prescription of the star
  formation rate with the molecule formation timescale;
\item depending on the star formation prescription, the best-fit break between
  regimes in the model is located near the transition region between
  the molecular-gas-dominated and atomic-gas-dominated parts of the
  galactic disk or closer to the optical radius;
\item
the viscous timescale is smaller than or comparable to the star formation timescale for galaxies less massive
than $10^{10}$~M$_{\odot}$, whereas it is much higher for the massive galaxies;
\item
as a consequence less massive galaxies can balance the gas loss due to star formation
by radial gas inflow within the galactic disk. In this way these galaxies can even
use the gas reservoir outside the optical radius. This is impossible for massive galaxies.
The star formation rate of massive galaxies is determined by the external gas mass accretion rate.
Massive galaxies depend thus on external infall with an angular momentum close
to that of the disk, whereas less massive galaxies can use large angular momentum gas 
located beyond the optical radius for star formation. 
\end{enumerate}

\begin{appendix}

\section{Model equations \label{sec:modelequations}}

\subsection{Star formation recipe according to \citet{Krumholz} (Eq.~\ref{eq:starform}) \label{sec:app1}}

The following equations are appropriate for the {\em inner disk} regime
where the local free fall time is the limiting timescale for star formation
and  $\Sigma_{*} \gg \Sigma$. For the actual numbers we assume 
$\dot{M}=0.2$~M$_{\odot}$yr$^{-1}$, $Q=2$, $\Sigma_{*}=100$~M$_{\odot}$yr$^{-1}$,
$\delta=2$, $\eta=7 \times 10^{-3}$, $\alpha=7.2 \times 10^{7} \times \big( \log(\frac{\Sigma_{*}+\Sigma}{\Sigma})
\big)^{-1}\ {\rm yr\,M_{\odot}pc^{-3}}=2 \times 10^{7}$~yr\,M$_{\odot}$pc$^{-3}$,
$\gamma=0.2$, and $\Omega=1.8 \times 10^{-8}$~yr$^{-1}$.

\begin{equation}
v_{\rm turb}=1.08\,G^{\frac{1}{3}}\dot{M}^{\frac{1}{6}}Q^{\frac{1}{6}}\delta^{\frac{1}{6}}
\xi^{\frac{1}{6}}\eta^{\frac{1}{6}}\Sigma_{*}^{\frac{1}{6}} \Omega^{-\frac{1}{6}}=19~{\rm km\,s}^{-1}\ ,
\label{eq:vturbsgz3}
\end{equation}
\begin{equation}
\phi_{\rm V}=0.50\,G^{\frac{2}{3}}\dot{M}^{\frac{5}{6}}\delta^{-\frac{19}{6}}Q^{-\frac{1}{6}}
\xi^{-\frac{7}{6}}\eta^{-\frac{7}{6}}\Sigma_{*}^{-\frac{1}{6}}\Omega^{\frac{1}{6}}=0.01\ ,
\label{eq:phiv3}
\end{equation}
\begin{equation}
\Sigma=0.50\,G^{-\frac{2}{3}}\dot{M}^{\frac{1}{6}}Q^{-\frac{5}{6}}\delta^{\frac{1}{6}}
\xi^{\frac{1}{6}}\eta^{\frac{1}{6}}\Sigma_{*}^{\frac{1}{6}}\Omega^{\frac{5}{6}}=16~{\rm M_{\odot}pc^{-2}}\ ,
\label{eq:sigmasgz3}
\end{equation}
\begin{equation} \label{eq:starform3}
\dot{\Sigma}_{*}=\eta \delta \Sigma_{*} Q^{-1} \Omega=1.3 \times 10^{-8}~{\rm M_{\odot}yr^{-1}pc^{-2}}\ ,
\end{equation}
\begin{equation}
l_{\rm driv}=0.43\,G^{\frac{1}{3}}\dot{M}^{\frac{2}{3}}Q^{\frac{2}{3}}\delta^{-\frac{1}{3}}
\xi^{-\frac{1}{3}}\eta^{-\frac{1}{3}}\Sigma_{*}^{-\frac{1}{3}}\Omega^{-\frac{2}{3}}=140~{\rm pc}\ ,
\label{eq:ldriv3}
\end{equation}
\begin{equation}
\nu=0.46\,G^{\frac{2}{3}}\dot{M}^{\frac{5}{6}}Q^{\frac{5}{6}}\delta^{-\frac{1}{6}}\xi^{-\frac{1}{6}}
\eta^{-\frac{1}{6}}\Sigma_{*}^{-\frac{1}{6}}\Omega^{-\frac{5}{6}}=2.8 \times 10^{-3}~{\rm pc^{2}yr^{-1}}\ ,
\label{eq:nu3}
\end{equation}
\begin{equation} \label{eq:sfe3}
SFE=1.99\,G^{\frac{2}{3}}\dot{M}^{-\frac{1}{6}}Q^{-\frac{1}{6}}\delta^{\frac{5}{6}}\xi^{-\frac{1}{6}}
\eta^{\frac{5}{6}}\Sigma_{*}^{\frac{5}{6}}\Omega^{\frac{1}{6}}=7.7 \times 10^{-10}~{\rm yr^{-1}}\ ,
\end{equation}
\begin{equation} \label{eq:fmol3}
f_{\rm mol}=0.74\,G^{-1}\dot{M}^{-\frac{1}{2}}Q^{-\frac{1}{2}}\delta^{\frac{3}{2}}
\xi^{\frac{1}{2}}\eta^{\frac{1}{2}}\Sigma_{*}^{\frac{1}{2}}\Omega^{\frac{1}{2}} \alpha^{-1}=0.80\ .
\end{equation}

The following equations are appropriate for the {\em inner disk} regime
where the local free fall time is the limiting timescale for star formation
and  $\Sigma \gg \Sigma_{*}$:

\begin{equation}
v_{\rm turb}=0.87\,G^{\frac{1}{5}}\dot{M}^{\frac{1}{5}}\delta^{\frac{1}{5}}
\xi^{\frac{1}{5}}\eta^{\frac{1}{5}}=12~{\rm km\,s}^{-1}\ ,
\label{eq:vturbsgz}
\end{equation}
\begin{equation}
\phi_{\rm V}=0.62\,G^{\frac{4}{5}}\dot{M}^{\frac{4}{5}}\delta^{-\frac{16}{5}}
\xi^{-\frac{6}{5}}\eta^{-\frac{6}{5}}=0.02\ ,
\label{eq:phiv}
\end{equation}
\begin{equation}
\Sigma=0.28\,G^{-\frac{4}{5}}\dot{M}^{\frac{1}{5}}Q^{-1}\delta^{\frac{1}{5}}
\xi^{\frac{1}{5}}\eta^{\frac{1}{5}}\Omega=7~{\rm M_{\odot}pc^{-2}}\ ,
\label{eq:sigmasgz1}
\end{equation}
\begin{equation} \label{eq:starform1}
\dot{\Sigma}_{*}=0.28\,G^{-\frac{4}{5}}\dot{M}^{\frac{1}{5}}Q^{-2}\delta^{\frac{6}{5}}\xi^{\frac{1}{5}}\eta^{\frac{6}{5}}\Omega^{2}
=9.2 \times 10^{-10}~{\rm M_{\odot}yr^{-1}pc^{-2}}\ ,
\end{equation}
\begin{equation}
l_{\rm driv}=0.66\,G^{\frac{3}{5}}\dot{M}^{\frac{3}{5}}Q\delta^{-\frac{2}{5}}
\xi^{-\frac{2}{5}}\eta^{-\frac{2}{5}}\Omega^{-1}=350~{\rm pc}\ ,
\label{eq:ldriv}
\end{equation}
\begin{equation}
\nu=0.57\,G^{\frac{4}{5}}\dot{M}^{\frac{4}{5}}Q\delta^{-\frac{1}{5}}\xi^{-\frac{1}{5}}
\eta^{-\frac{1}{5}}\Omega^{-1}=6.0 \times 10^{-4}~{\rm pc^{2}yr^{-1}}\ ,
\label{eq:nu}
\end{equation}
\begin{equation} \label{eq:sfe}
SFE=\eta \delta Q^{-1} \Omega=1.3 \times 10^{-10}~{\rm yr^{-1}}\ ,
\end{equation}
\begin{equation} \label{eq:fmol}
f_{\rm mol}=0.39\,G^{-\frac{7}{5}}\dot{M}^{-\frac{2}{5}}Q^{-1}\delta^{\frac{8}{5}}
\xi^{\frac{3}{5}}\eta^{\frac{3}{5}}\Omega \alpha^{-1}=0.21\ .
\end{equation}

Note that the turbulent velocity and the volume filling factor
are constant and thus independent of the galactocentric radius.

The following equations are appropriate for a selfgravitating {\em outer disk} where the molecular
formation timescale is the relevant timescale for star formation and $\Sigma \gg \Sigma_{*}$:

\begin{equation}
v_{\rm turb}=0.54\,G^{-\frac{1}{2}}Q^{-\frac{1}{2}} \delta \xi^{\frac{1}{2}}\eta^{\frac{1}{2}}
(\gamma \alpha)^{-\frac{1}{2}} \Omega^{\frac{1}{2}}=12~{\rm km\,s}^{-1}\ ,
\label{eq:vturbsgz1}
\end{equation}
\begin{equation}
\phi_{\rm V}=10.6\,G^{5}\dot{M}^{2}Q^{3}\delta^{-8} \xi^{-3} \eta^{-3} \alpha^{3} \Omega^{-3}=0.01\ ,
\label{eq:phiv1}
\end{equation}
\begin{equation}
\Sigma=0.17\,G^{-\frac{3}{2}} Q^{-\frac{3}{2}} \delta \xi^{\frac{1}{2}} \eta^{\frac{1}{2}} 
\alpha^{-\frac{1}{2}} \Omega^{\frac{3}{2}}=7~{\rm M_{\odot}pc^{-2}}\ ,
\label{eq:sigmasgz2}
\end{equation}
\begin{equation} \label{eq:starform2}
\dot{\Sigma}_{*}=0.02\,G^{-5}\dot{M}^{-1}Q^{-5} \delta^{6} \xi^{2} \eta^{3} \alpha^{-3} \Omega^{5}=
1.4 \times 10^{-9}~{\rm M_{\odot}yr^{-1}pc^{-2}}\ ,
\end{equation}
\begin{equation}
l_{\rm driv}=1.7\,G^{2}\dot{M} Q^{2} \delta^{-2} \xi^{-1} \eta^{-1} \alpha \Omega^{-2}=330~{\rm pc}\ ,
\label{eq:ldriv1}
\end{equation}
\begin{equation}
\nu=0.92\,G^{\frac{3}{2}}\dot{M} Q^{\frac{3}{2}} \delta^{-\frac{3}{2}} \xi^{-\frac{1}{2}} \eta^{-\frac{1}{2}}
\alpha^{\frac{1}{2}}\Omega^{-\frac{3}{2}}=3.0 \times 10^{-3}~{\rm pc^{2}yr^{-1}}\ ,
\label{eq:nu1}
\end{equation}
\begin{equation} \label{eq:sfe1}
SFE=0.09\,G^{-\frac{7}{2}}\dot{M}^{-1}Q^{-\frac{7}{2}} \delta^{5}
\xi^{\frac{3}{2}} \eta^{\frac{5}{2}} (\gamma \alpha)^{-\frac{5}{2}}\Omega^{\frac{7}{2}}=1.4 \times 10^{-10}~{\rm yr^{-1}}\ ,
\end{equation}
\begin{equation} \label{eq:fmol1}
f_{\rm mol}=0.09\,G^{-\frac{7}{2}} \dot{M}^{-1} Q^{-\frac{5}{2}} \delta^{4}
\xi^{\frac{3}{2}} \eta^{\frac{3}{2}} \gamma^{-\frac{3}{2}} \alpha^{-\frac{5}{2}}\Omega^{\frac{5}{2}}=0.23\ .
\end{equation}

\subsection{Star formation according to VB03 (Eq.~\ref{eq:starformm})}

The following equations are appropriate for the {\em inner disk} regime
where the local free fall time is the limiting timescale for star formation
and $\Sigma_{*} \gg \Sigma$. For the actual numbers we assume 
$\dot{M}=0.2$~M$_{\odot}$yr$^{-1}$, $Q=2$, $\Sigma_{*}=100$~M$_{\odot}$yr$^{-1}$,
$\delta=2$, $\alpha=7.2 \times 10^{7} \times \big( \log(\frac{\Sigma_{*}+\Sigma}{\Sigma})
\big)^{-1}\ {\rm yr\,M_{\odot}pc^{-3}}=2 \times 10^{7}$~yr\,M$_{\odot}$pc$^{-3}$,
$\gamma=0.2$, and $\Omega=1.8 \times 10^{-8}$~yr$^{-1}$.

\begin{equation}
v_{\rm turb}=1.02\,G^{\frac{5}{13}}\dot{M}^{\frac{3}{13}}Q^{\frac{2}{13}}\delta^{-\frac{1}{13}}
\xi^{\frac{1}{13}}\Sigma_{*}^{\frac{2}{13}} \Omega^{-\frac{2}{13}}=19~{\rm km\,s}^{-1}\ ,
\label{eq:vturbsgz4}
\end{equation}
\begin{equation}
\phi_{\rm V}=0.73\,G^{\frac{4}{13}}\dot{M}^{\frac{5}{13}}\delta^{-\frac{19}{13}}Q^{-\frac{1}{13}}
\xi^{-\frac{7}{13}}\Sigma_{*}^{-\frac{1}{13}}\Omega^{\frac{1}{13}}=9 \times 10^{-3}\ ,
\label{eq:phiv4}
\end{equation}
\begin{equation}
\Sigma=0.33\,G^{-\frac{8}{13}}\dot{M}^{\frac{3}{13}}Q^{-\frac{11}{13}}\delta^{-\frac{1}{13}}
\xi^{\frac{1}{13}}\Sigma_{*}^{\frac{2}{13}}\Omega^{\frac{11}{13}}=12~{\rm M_{\odot}pc^{-2}}\ ,
\label{eq:sigmasgz4}
\end{equation}
\begin{equation} \label{eq:starform4}
\dot{\Sigma}_{*}= 0.73\,G^{\frac{4}{13}}\dot{M}^{\frac{5}{13}}Q^{-\frac{14}{13}}\delta^{-\frac{6}{13}}
\xi^{-\frac{7}{13}}\Sigma_{*}^{\frac{12}{13}}\Omega^{\frac{14}{13}}=
1.5 \times 10^{-8}~{\rm M_{\odot}yr^{-1}pc^{-2}}\ ,
\end{equation}
\begin{equation}
l_{\rm driv}=0.48\,G^{\frac{3}{13}}\dot{M}^{\frac{7}{13}}Q^{\frac{9}{13}}\delta^{\frac{2}{13}}
\xi^{-\frac{2}{13}}\Sigma_{*}^{-\frac{4}{13}}\Omega^{-\frac{9}{13}}=140~{\rm pc}\ ,
\label{eq:ldriv4}
\end{equation}
\begin{equation}
\nu=0.49\,G^{\frac{8}{13}}\dot{M}^{\frac{10}{13}}Q^{\frac{11}{13}}\delta^{\frac{1}{13}}\xi^{-\frac{1}{13}}
\Sigma_{*}^{-\frac{2}{13}}\Omega^{-\frac{11}{13}}=2.8 \times 10^{-3}~{\rm pc^{2}yr^{-1}}\ ,
\label{eq:nu4}
\end{equation}
\begin{equation} \label{eq:sfe4}
SFE=2.23\,G^{\frac{12}{13}}\dot{M}^{\frac{2}{13}}Q^{-\frac{3}{13}}\delta^{-\frac{5}{13}}\xi^{-\frac{8}{13}}
\Sigma_{*}^{\frac{10}{13}}\Omega^{\frac{3}{13}}=1.3 \times 10^{-9}~{\rm yr^{-1}}\ ,
\end{equation}
\begin{equation} \label{eq:fmol4}
f_{\rm mol}=0.63\,G^{-\frac{11}{13}}\dot{M}^{-\frac{4}{13}}Q^{-\frac{7}{13}}\delta^{\frac{10}{13}}
\xi^{\frac{3}{13}}\Sigma_{*}^{\frac{6}{13}}\Omega^{\frac{7}{13}} \alpha^{-1}=0.88\ .
\end{equation}

The following equations are appropriate for the {\em inner disk} regime
where the local free fall time is the limiting timescale for star formation
and $\Sigma \gg \Sigma_{*}$:

\begin{equation}
v_{\rm turb}=0.82\,G^{\frac{3}{11}}\dot{M}^{\frac{3}{11}}\delta^{-\frac{1}{11}}
\xi^{\frac{1}{11}}=13~{\rm km\,s}^{-1}\ ,
\label{eq:vturbsgz5}
\end{equation}
\begin{equation}
\phi_{\rm V}=0.81\,G^{\frac{4}{11}}\dot{M}^{\frac{4}{11}}\delta^{-\frac{16}{11}}
\xi^{-\frac{6}{11}}=0.01\ ,
\label{eq:phiv5}
\end{equation}
\begin{equation}
\Sigma=0.26\,G^{-\frac{8}{11}}\dot{M}^{\frac{3}{11}}Q^{-1}\delta^{-\frac{1}{11}}
\xi^{\frac{1}{11}}\Omega=8~{\rm M_{\odot}pc^{-2}}\ ,
\label{eq:sigmasgz5}
\end{equation}
\begin{equation} \label{eq:starform5}
\dot{\Sigma}_{*}=0.21\,G^{-\frac{4}{11}}\dot{M}^{\frac{7}{11}}Q^{-2}\delta^{-\frac{6}{11}}\xi^{-\frac{5}{11}}\Omega^{2}=
1.4 \times 10^{-9}~{\rm M_{\odot}yr^{-1}pc^{-2}}\ ,
\end{equation}
\begin{equation}
l_{\rm driv}=0.71\,G^{\frac{5}{11}}\dot{M}^{\frac{5}{11}}Q\delta^{\frac{2}{11}}
\xi^{-\frac{2}{11}}\Omega^{-1}=290~{\rm pc}\ ,
\label{eq:ldriv5}
\end{equation}
\begin{equation}
\nu=0.61\,G^{\frac{8}{11}}\dot{M}^{\frac{8}{11}}Q\delta^{\frac{1}{11}}\xi^{-\frac{1}{11}}\Omega^{-1}=
4.1 \times 10^{-3}~{\rm pc^{2}yr^{-1}}\ ,
\label{eq:nu5}
\end{equation}
\begin{equation} \label{eq:sfe5}
SFE=0.81\,G^{\frac{4}{11}}\dot{M}^{\frac{4}{11}}Q^{-1}\delta^{-\frac{5}{11}}\xi^{-\frac{6}{11}}\Omega=
1.9 \times 10^{-10}~{\rm yr^{-1}}\ ,
\end{equation}
\begin{equation} \label{eq:fmol5}
f_{\rm mol}=0.34\,G^{-\frac{13}{11}}\dot{M}^{-\frac{2}{11}}Q^{-1}\delta^{\frac{8}{11}}
\xi^{\frac{3}{11}}\Omega \alpha^{-1}=0.27\ .
\end{equation}

Note that the turbulent velocity and the volume filling factor
are constant and thus independent of the galactocentric radius.

The following equations are appropriate for a selfgravitating {\em outer disk} where the molecular
formation timescale is the relevant timescale for star formation and $\Sigma \gg \Sigma_{*}$:

\begin{equation}
v_{\rm turb}=0.71\,G^{\frac{1}{8}}\dot{M}^{\frac{1}{4}}Q^{-\frac{1}{8}}\xi^{\frac{1}{8}}
(\gamma \alpha)^{-\frac{1}{8}} \Omega^{\frac{1}{8}}=13~{\rm km\,s}^{-1}\ ,
\label{eq:vturbsgz6}
\end{equation}
\begin{equation}
\phi_{\rm V}=1.95\,G^{\frac{5}{4}}\dot{M}^{\frac{1}{2}}Q^{\frac{3}{4}}\delta^{-2}
\xi^{-\frac{3}{4}} (\gamma \alpha)^{\frac{3}{4}} \Omega^{-\frac{3}{4}}=9 \times 10^{-3}\ ,
\label{eq:phiv6}
\end{equation}
\begin{equation}
\Sigma=0.23\,G^{-\frac{7}{8}}\dot{M}^{\frac{1}{4}}Q^{-\frac{9}{8}}\xi^{\frac{1}{8}} 
(\gamma \alpha)^{-\frac{1}{8}} \Omega^{\frac{9}{8}}=8~{\rm M_{\odot}pc^{-2}}\ ,
\label{eq:sigmasgz6}
\end{equation}
\begin{equation} \label{eq:starform6}
\dot{\Sigma}_{*}=0.09\,G^{-\frac{5}{4}}\dot{M}^{\frac{1}{2}}Q^{-\frac{11}{4}}\xi^{-\frac{1}{4}} 
(\gamma \alpha)^{-\frac{3}{4}} \Omega^{\frac{11}{4}}=1.7 \times 10^{-9}~{\rm M_{\odot}yr^{-1}pc^{-2}}\ ,
\end{equation}
\begin{equation}
l_{\rm driv}=0.93\,G^{\frac{3}{4}}\dot{M}^{\frac{1}{2}}Q^{\frac{5}{4}}\xi^{-\frac{1}{4}}
(\gamma \alpha)^{\frac{1}{4}}\Omega^{-\frac{5}{4}}=270~{\rm pc}\ ,
\label{eq:ldriv6}
\end{equation}
\begin{equation}
\nu=0.69\,G^{\frac{7}{8}}\dot{M}^{\frac{3}{4}}Q^{\frac{9}{8}} \xi^{-\frac{1}{8}}
(\gamma \alpha)^{\frac{1}{8}}\Omega^{-\frac{9}{8}}=
3.9 \times 10^{-3}~{\rm pc^{2}yr^{-1}}\ ,
\label{eq:nu6}
\end{equation}
\begin{equation} \label{eq:sfe6}
SFE=0.39\,G^{-\frac{3}{8}}\dot{M}^{\frac{1}{4}}Q^{-\frac{13}{8}}
\xi^{-\frac{3}{8}}(\gamma \alpha)^{-\frac{5}{8}}\Omega^{\frac{13}{8}}=
2.1 \times 10^{-10}~{\rm yr^{-1}}\ ,
\end{equation}
\begin{equation} \label{eq:fmol6}
f_{\rm mol}=0.22\,G^{-\frac{13}{8}} \dot{M}^{-\frac{1}{4}} Q^{-\frac{11}{8}} \delta
\xi^{\frac{3}{8}} \gamma^{-\frac{3}{8}} \alpha^{-\frac{11}{8}}\Omega^{\frac{11}{8}}=0.29\ .
\end{equation}

\subsection{Critical angular velocity}

Here we give the expressions for the critical angular velocity $\Omega_{\rm crit}$ where
$t^{\rm l}_{\rm ff}= \gamma t^{\rm l}_{\rm mol}$ ($\Sigma \gg \Sigma_{*}$). 
For the actual numbers we assume 
$\dot{M}=0.2$~M$_{\odot}$yr$^{-1}$, $Q=2$, $\Sigma_{*}=100$~M$_{\odot}$yr$^{-1}$,
$\delta=2$, $\eta=7 \times 10^{-3}$, $\alpha=7.2 \times 10^{7} \times \big( \log(\frac{\Sigma_{*}+\Sigma}{\Sigma})
\big)^{-1}\ {\rm yr\,M_{\odot}pc^{-3}}=2 \times 10^{7}$~yr\,M$_{\odot}$pc$^{-3}$,
$\gamma=0.2$, and $\Omega=1.8 \times 10^{-8}$~yr$^{-1}$.

Star formation recipe according to \citet{Krumholz} (Eq.~\ref{eq:starform}):

\begin{equation}
\Omega_{\rm crit}=2.6\,G^{\frac{7}{5}}\dot{M}^{\frac{2}{5}}\delta^{-\frac{8}{5}}\xi^{-\frac{3}{5}}\eta^{-\frac{3}{5}}\gamma \alpha Q
=1.7 \times 10^{-8}~{\rm yr^{-1}}\ .
\end{equation}

Star formation according to VB03 (Eq.~\ref{eq:starformm}):

\begin{equation}
\Omega_{\rm crit}=2.9\,G^{\frac{13}{11}}\dot{M}^{\frac{2}{11}}\delta^{-\frac{8}{11}}\xi^{-\frac{3}{11}}\gamma \alpha Q=
1.3 \times 10^{-8}~{\rm yr^{-1}}\ .
\end{equation}

Assuming a constant rotation velocity of $v_{\rm rot}=200$~km\,s$^{-1}$, we obtain critical radii of
12.4~kpc and 16.2~kpc, respectively.

\end{appendix}




\begin{table*}
\begin{center}
\caption{Model Parameters.\label{tab:parameters}}
\begin{tabular}{lll}
\tableline\tableline
Parameter & Unit & Explanation \\
\tableline
$G=5 \times 10^{-15}$ & pc$^{3}$yr$^{-1}$M$_{\odot} ^{-1}$ & gravitation constant \\
$\kappa$ & yr$^{-1}$ & epicyclic frequency \\
$Q$ & & Toomre parameter \\
$R$ & pc &galactocentric radius \\
$H$ & pc & thickness of the gas disk\\
$H_{*}$ & pc & thickness of the stellar disk \\
$R_{\rm break}$ & pc & break radius between star formation regimes \\
$l_{\rm cl}$ & pc & cloud size \\
$v_{\rm rot}$ & pc\,yr$^{-1}$ & rotation velocity \\
$\Omega=v_{\rm rot}/R$ & yr$^{-1}$ & angular velocity \\
$\Phi_{\rm V}$ & & volume filling factor \\
$\Phi_{\rm A}=\Phi_{\rm V}\,H/l_{\rm cl}$ & & area filling factor \\
$\rho$ & M$_{\odot}$pc$^{-3}$ & disk midplane gas density\\
$\rho_{\rm cl}=\rho/\Phi_{\rm V}$ & M$_{\odot}$pc$^{-3}$ & cloud density \\
$\dot{\rho}_{*}$ & M$_{\odot}$pc$^{-3}$yr$^{-1}$ & star formation rate \\
$\Sigma$ & M$_{\odot}$pc$^{-2}$ & gas surface density \\
$\Sigma_{*}$ & M$_{\odot}$pc$^{-2}$ & stellar surface density \\
$\dot{\Sigma}_{*}$ & M$_{\odot}$pc$^{-2}$yr$^{-1}$ & star formation rate \\
$\xi=4.6 \times 10^{-8}$ & pc$^2$yr$^{-2}$ & constant relating SN energy input to SF \\
$\dot{M}$ & M$_{\odot}$yr$^{-1}$ & disk mass accretion rate \\
$v_{\rm turb}$ & pc\,yr$^{-1}$ & gas turbulent velocity dispersion \\
$v_{\rm rad}$ & pc\,yr$^{-1}$ & gas radial velocity \\
$v_{\rm disp}^{*}$ & pc\,yr$^{-1}$ & stellar vertical velocity dispersion \\
$\nu$ & pc$^{2}$yr$^{-1}$ & viscosity \\
$f_{\rm mol}=\Sigma_{\rm H_{2}}/(\Sigma_{\rm HI}+\Sigma_{\rm H_{2}})$ &  & molecular fraction \\
$\alpha$ & yr\,M$_{\odot}$pc$^{-3}$ & constant of molecule formation timescale \\
$l_{\rm driv}$ & pc & turbulent driving length scale \\
$l_{\rm diss}$ & pc & turbulent dissipation length scale \\
$\delta$ & & scaling between driving and dissipation length scale \\
$SFE=\dot{\Sigma}_{*}/\Sigma$ & yr$^{-1}$ & star formation efficiency \\
$t_{\rm ff}^{l}$ & yr & cloud free fall timescale at size $l$ \\
$t_{\rm turb}^{l}$ & yr & cloud turbulent timescale at size $l$ \\
$t_{\rm mol}^{l}$ & yr & cloud molecule formation timescale at size $l$ \\
$\eta$ & & star formation efficiency per free fall time \\
$\gamma$ & & $t_{\rm sf}^{l}= \gamma t_{\rm mol}^{l}$ at the break radius \\
\tableline
\end{tabular}
\end{center}
\end{table*}

\begin{table*}
\begin{center}
\caption{Properties of Sample Galaxies (from Leroy et al. 2008).\label{tab:galaxies}}
\begin{tabular}{lccccccccc}
\tableline\tableline
Galaxy & Dist. & i & P.A. & Morph. & $M_{\rm B}$ & $R_{\rm 25}$ & log $M_{*}$ & SFR &  $l_{*}$  \\
 & (Mpc) & ($^{\circ}$) & ($^{\circ}$) &  & (mag) & (kpc) & (M$_{\odot}$) & (M$_{\odot}$yr$^{-1}$) &  (kpc)  \\
\tableline
IC 2574 &  4.0 &  53 &  56 & Irr & -18.0 &  7.5 &  8.7 &   0.07 &   2.1 \\
NGC 4214 &  2.9 &  44 &  65 & Irr & -17.4 &  2.9 &  8.8 &   0.11 &   0.7 \\
NGC 2976 &  3.6 &  65 & 335 & Sc & -17.8 &  3.8 &    9.1 &   0.09 &  0.9 \\
NGC 7793 &  3.9 &  50 & 290 & Scd & -18.7 &  6.0 &   9.5 &   0.24 &  1.3 \\
NGC 2403 &  3.2 &  63 & 124 & SBc & -19.4 &  7.3  &  9.7 &   0.38 &  1.6 \\
NGC 0925 &  9.2 &  66 & 287 & SBcd & -20.0 & 14.2 &   9.9 &   0.56 &  4.1 \\
NGC 0628 &  7.3 &   7 &  20 & Sc & -20.0 & 10.4 & 10.1 &   0.81 &  2.3 \\
NGC 3198 & 13.8 &  72 & 215 & SBc & -20.7 & 13.0 &  10.1 &  0.93 &  3.2 \\
NGC 3184 & 11.1 &  16 & 179 & SBc & -19.9 & 11.9 &  10.3 &  0.90 &  2.4 \\
NGC 4736 &  4.7 &  41 & 296 & Sab & -20.0 &  5.3 &  10.3 &   0.48 &  1.1 \\
NGC 3351 & 10.1 &  41 & 192 & SBb & -19.7 & 10.6 &  10.4 &   0.94 &  2.2 \\
NGC 6946 &  5.9 &  33 & 243 & SBc & -20.9 &  9.8 &  10.5 &   3.24 &  2.5 \\
NGC 3627 &  9.3 &  62 & 173 & SBb & -20.8 & 13.9 &  10.6 &   2.22 &  2.8 \\
NGC 5194 &  8.0 &  20 & 172 & SBc & -21.1 &  9.0 &  10.6 &   3.13 &   2.8 \\
NGC 3521 & 10.7 &  73 & 340 & SBbc & -20.9 & 12.9 &  10.7 &  2.10 &  2.9 \\
NGC 2841 & 14.1 &  74 & 153 & Sb & -21.2 & 14.2 &  10.8 &  0.74 &   4.0 \\
NGC 5055 & 10.1 &  59 & 102 & Sbc & -20.6 & 17.4 &  10.8 &  2.12 &  3.2 \\
NGC 7331 & 14.7 &  76 & 168 & SAb & -21.7 & 19.6 &  10.9 & 2.99 &  3.3 \\
\tableline
\end{tabular}
\end{center}
\end{table*}

\begin{table*}
\begin{center}
\caption{Model results for the KM05 and VB03 star formation prescriptions. \label{tab:galaxies1}}
\begin{tabular}{lccccccccc}
\tableline\tableline
Galaxy & $\dot{M}$ & $\delta$ & $\gamma$ & $R_{\rm break}$ & $c_{\rm H_2}$ & $c_{\rm v_{\rm disp}}$ & $c_{\rm SFE}$ & $g$ & $< t_{\rm ff} >$ \\
 & (M$_{\odot}$yr$^{-1}$) &  &  & (kpc) & & & & & (Myr) \\
\tableline
KM05 & & & & & & & & \\
\hline
IC2574   &   0.22   &   2.0  &    0.07   &   1.1   &  0.63   &   0.79   &   0.71 &    0.98 & 29.6 \\
NGC4214  &  0.08  &    1.3  &   0.11   &   1.3 &    0.63  &   0.79  &   0.71 &    0.15 & 9.5 \\
NGC2976  &   0.31  &    3.2  &    -   &   -  &   0.63  &    1.00   &   1.00  &    2.59 & 4.1 \\
NGC7793  &  0.06  &    1.4  &   0.11    &  3.9   &  0.63  &   0.79   &  0.71 &    0.69 & 6.0 \\
NGC2403  &   0.22   &   1.8  &  0.07   &   4.9  &    1.00   &   1.00  &   0.71  &   0.31 & 8.8 \\
NGC0925   &  0.11  &    2.0 &   0.07   &   9.1   &  0.63  &   0.79  &   0.71  &   0.97 & 12.5 \\
NGC0628  &  0.08  &    1.6  &  0.13   &   7.6   &  0.63    &  1.00   &   1.00 &    0.95 & 4.7 \\
NGC3198   &  0.11  &    1.8   &  0.07  &    11.0   &   0.63  &   0.79  &    1.00  &   0.94 & 7.2 \\
NGC3184  &   0.11  &    2.0 &  0.10    &  10.0  &   0.63   &   1.00   &   1.41  &    1.24 & 3.8 \\
NGC4736  &    0.89  &    2.0  &    -   &  -   &  0.63   &   1.26   &  1.41   &   2.34 & 7.7 \\
NGC3351  &  0.06  &    1.6  &   0.10  &    8.6 &    0.63   &   1.00  &   0.71  &   0.79 & 5.8 \\
NGC6946  &   0.32   &   2.5 &     -   &   -   &  0.63  &    1.00   &   1.00  &    1.47 & 3.0 \\
NGC3627  &   0.22  &    2.0   &   -   &   -   &  0.63  &   0.79  &   1.00  &   0.47 & 4.4 \\
NGC5194  &    1.26  &    1.0  &    -   &   -  &   0.63  &   0.79  &   0.71   &   3.29 & 22.1 \\
NGC3521   &  0.22  &    1.8  &  -   &   -   &  0.63  &   0.79   &   1.41  &    5.82 & 4.9 \\
NGC2841   &  0.08   &   2.0  &   0.17   &   11.3  &   0.63  &   0.79   &  0.71  &   0.42 & 6.0 \\
NGC5055   &  0.32  &    2.0  &  0.07   &   12.5  &   0.63  &   0.79  &    1.41   &   7.83 & 5.9 \\
NGC7331   &  0.22   &   2.0 &   0.09   &   13.9  &   0.63   &  0.79    &  1.41  &    1.37 & 4.4 \\
\hline
VB03 & & & & & & & & \\
\hline
IC 2574   &  0.16  &    8.9   &  1.00  &   0.00   &  0.63  &   0.79  &   0.71  &   0.21 & 5.2 \\
NGC 4214  &   0.06  &    1.0   &  1.00   &  0.00  &   0.63  &   0.79   &  0.71  &   0.21 & 11.2 \\
NGC 2976  &   0.32  &    3.5  &    -  &   -  &   0.63   &   1.00   &   0.71   &   2.44 & 4.1 \\
NGC 7793  &  0.16  &    1.0  &   1.00   &  0.00  &   0.63  &   1.00  &   0.71  &   0.33 & 12.5 \\
NGC 2403  &   0.22   &   1.3  &   1.00   &   0.00   &   0.63   &   1.00   &   0.71   &  0.11 & 14.2 \\
NGC 0925  &   0.16    &  1.0   &  0.33   &  0.00  &   0.63   &  0.79   &  0.71   &  0.41 & 32.0 \\
NGC 0628  &   0.08  &    1.8   &   1.00   &  0.00  &   0.63  &    1.00  &   1.00  &   1.01 & 4.6 \\
NGC 3198  &   0.11   &   2.5   &  0.33   &   4.3   &   1.00   &  0.79   &  1.00  &   0.93 & 5.1 \\
NGC 3184  &   0.16  &    4.0   &  0.33   &   6.7   &   1.58   &   1.00   &   1.41  &   1.31 & 2.3 \\
NGC 4736  &    0.63  &    7.9  &    -  &   -   &   1.58  &    1.26  &   1.41   &   2.59 & 1.7 \\
NGC 3351  &  0.06  &    1.8   &  0.33    &  2.7  &   0.63   &   1.00   &  0.71  &   0.84 & 5.1 \\
NGC 6946  &   0.45   &   7.9   &   -   &   -   &   1.58   &   1.00   &   1.00   &   1.13 & 1.2 \\
NGC 3627  &   0.32   &   8.9  &    -   &  -   &   1.58   &  0.79  &   0.71   &  0.40 & 1.3\\
NGC 5194  &   0.06   &   8.9   &   -   &  -  &    1.58  &   0.79  &   0.71    &  4.30 & 0.8 \\
NGC 3521 &    0.32   &   8.9  &    0.25   &   11.7   &   1.58  &   0.79   &   1.41  &    4.22 & 1.1 \\
NGC 2841  &   0.11   &   3.5  &   0.33  &   5.1   &   1.00  &   0.79  &   0.71  &   0.41 & 4.3 \\
NGC 5055  &   0.89   &  8.9   &  1.00  &    5.1   &   1.58   &   1.00   &   1.41   &   4.06 & 2.0 \\
NGC 7331   &  0.22   &   4.5   &  0.50  &    6.8   &   1.58  &   0.79  &    1.41  &   0.67 & 2.0 \\
\tableline
\end{tabular}
\end{center}
\end{table*}

\end{document}